\RequirePackage{fix-cm}
\documentclass[twocolumn,epjc3]{svjour3}  
\smartqed  
\RequirePackage{graphicx}
\RequirePackage{latexsym}
\RequirePackage[numbers,sort&compress]{natbib}
\RequirePackage[colorlinks,citecolor=blue,urlcolor=blue,linkcolor=blue]{hyperref}
\usepackage{amsmath}
\usepackage[dvipsnames, table]{xcolor}
\usepackage{float}
\usepackage{simpler-wick}
\usepackage{eqnarray}
\usepackage{listings}
\usepackage{makecell, cellspace}
\usepackage{multirow}
\usepackage{array}
\usepackage{booktabs}
\usepackage{subcaption}
\definecolor{dkgreen}{rgb}{0,0.6,0}
\definecolor{gray}{rgb}{0.05,0.05,0.05}
\definecolor{mauve}{rgb}{0.58,0,0.82}
\definecolor{Fuchsia}{rgb}{0.098,0.098,.439}
\definecolor{DarkViolet}{rgb}{0.4,0.0,.4}
\definecolor{PinkViolet}{rgb}{0.4,0.0,.2}
\definecolor{twilightlavender}{rgb}{0.54, 0.29, 0.42}
\definecolor{DarkBlue}{rgb}{0.0,0.4,0.4}
\definecolor{GreenBlue}{rgb}{0.0,0.2,0.4}
\definecolor{LightBlue}{rgb}{0.0,0.5,0.1}

\usepackage{graphicx}
\usepackage{dcolumn}
\usepackage{bm}

\journalname{}
\begin{document}

\title{Down type iso-singlet quarks at the HL-LHC and FCC-hh
}


\author{Arpon Paul\thanksref{e1,addr1}
        \and
        Sezen Sekmen\thanksref{e2,addr2}
        \and
        Gokhan Unel\thanksref{e3,addr3}
}

\thankstext{e1}{e-mail: \href{mailto:apaul@ictp.it}{apaul@ictp.it}}
\thankstext{e2}{e-mail: \href{mailto:ssekmen@cern.ch}{ssekmen@cern.ch}}
\thankstext{e3}{e-mail: \href{mailto:gokhan.unel@cern.ch}{gokhan.unel@cern.ch}}

\institute{The Abdus Salam International Centre for Theoretical Physics,  Strada Costiera 11,
I - 34151, Trieste, Italy. \label{addr1}
           \and
Kyungpook National University, Center for High Energy Physics, Daegu, South Korea.\label{addr2}
           \and
           University of California at Irvine, Physics Department, Irvine, CA 92697, USA.\label{addr3}
}

\date{}

\maketitle

\begin{abstract}
We study the discovery potential of down type iso-singlet quarks, $D$, predicted by the $E_6$ GUT model in the ${pp\rightarrow D\bar{D}\rightarrow ZZd\bar{d} \rightarrow \ell^+\ell^-\ell^+\ell^- d\bar{d}}$ channel, at the HL-LHC and FCC-hh colliders. The analysis is performed using a high level analysis description language and its runtime interpreter. The study shows that, using solely this channel, HL-LHC can discover $D$ quarks up to a mass of 730 GeV whereas FCC-hh up to 2980 GeV with data collected in their complete run periods.
\end{abstract}

\section{Introduction}
The long-awaited discovery of the Higgs boson at the LHC experiments \cite{20121} in the year 2012 completed the experimental validation of the standard model (SM). However, there are some well known issues that are not addressed by the SM, such as the mass hierarchy problem, the unification of the fundamental interactions, the origin of the baryon asymmetry of the Universe, and a plausible explanation for dark matter. To address these issues, SM is proposed to be extended into a more complete theory. In general, candidate extensions predict the existence of new fundamental particles and interactions. The Large Hadron Collider (LHC) experiments are conducting a great diversity of searches for discovering these new particles and interactions. Results of all these searches so far have been found to be consistent with the SM predictions. The forthcoming High Luminosity Large Hadron Collider (HL-LHC) and Future Circular Collider (FCC) machines, with their higher luminosity, energy and better detector acceptance and efficiency, will increase the sensitivity of these searches, and expand them to more difficult scenarios, enabling access to higher particle masses and lower effective cross sections. 

One class of candidate extensions to the SM consists of Grand Unified Theories (GUTs) based on a gauge group larger than that of the SM. The GUT models merge strong and electroweak interactions in a single gauge group, thereby allowing a solution to at least two of the above mentioned problems, namely, the complete unification of the fundamental interactions (except gravity) and the baryon asymmetry of the observed Universe. Specifically, when unifying gravity with other interactions both within the contexts of the superstring and supergravity theories, the exceptional Lie group $E_6$ has been shown to be the gauge symmetry group which can be compactified from 10 (or 11) dimensions down to the $3+1$ that we observe \cite{HEWETT1989193}.

The GUT model using the Exceptional Lie Group $E_6$ as the gauge symmetry group is referred to as the $E_6$ model. It predicts the existence of iso-singlet quarks (in literature, denoted by $D, S,$ and $B$) having charge $Q=-1/3$. The discovery potential of the ATLAS experiment for the down type iso-singlet quark $D$ of the first SM family  has  been previously investigated in~\cite{Mehdiyev2007,Mehdiyev2008}. The discovery reach for $D$ quarks were estimated at a phenomenology study before the LHC data taking to be 950 GeV for 100 fb$^{-1}$ integrated luminosity using the combination of all $D$ decay channels~\cite{Mehdiyev2008}.

Dedicated searches for down-type iso-singlet quarks predicted by the model described in this paper in the LHC data are currently ongoing in the ATLAS experiment. In the meanwhile, the closest estimates of sensitivity come from searches for vector-like quarks (VLQs), which have similar production mechanisms. However almost all existing VLQ searches are exclusively designed to target third generation vector-like partners $B$ and $T$ of the bottom and top quarks. The most stringent limits to date come from an ATLAS combination of 7 VLQ searches performed with 13 TeV data, looking at different final states~\cite{Aaboud:2018xuw,Aaboud:2017zfn,Aaboud:2018uek,Aaboud:2017qpr,Aaboud:2018saj,Aaboud:2018xpj,Aaboud:2018wxv}, which excluded $T$ ($B$) masses below 1.31 (1.03) TeV for any combination of decays into SM  particles~\cite{Aaboud:2018pii}. Several CMS studies also searched for third generation VLQs.  One search in the single lepton channel excluded $T$ masses less than 1.295 TeV in exclusive decays to $tW$~\cite{Sirunyan:2017pks}. A different fully hadronic search excluded $T$ and $B$ quark masses between 0.74-1.37 TeV~\cite{Sirunyan:2019sza}, while a leptonic search excluded $T$ quarks with masses below 1.14 to 1.30 TeV and $B$ quarks with masses below 0.91 to 1.24 TeV~\cite{Sirunyan:2018omb} for various branching fraction combinations.  However those limits do not directly apply to the down-type isosinglet quark scenario studied here, as the searches mainly focus on third generation final states that contain $b$ quarks.  For the specific case of light-flavor VLQ, a CMS search with at least one lepton excluded pair-produced VLQs below masses 845 and 685 GeV for branching ratios $B(W) = 1$ and $B(W) = 0.5$, $B(Z) = B(H) = 0.25$, respectively~\cite{Sirunyan:2017lzl}. 

In this study, we investigate the possibility of observing the pair production of first generation down type iso-singlet quarks $D$, in the decay channel $D \rightarrow Zd \rightarrow \ell^+\ell^-d$ (where $\ell = e, \mu$) using the 4 leptons plus 2 jets final state at the HL-LHC and the proton-proton scenario for the FCC. Due to its low effective cross section, this process could not be observed at the current LHC conditions.  With their higher luminosity, energy and detector acceptances, HL-LHC and FCC are expected to significantly improve sensitivity in this channel. Despite its low effective cross section, exploring this channel is critical, as it provide the most precise reconstruction of the $D$ quark mass in case of discovery.

Additionally, this work aims to test the feasibility of a new and practical analysis writing approach for high energy physics.  The search method in this study is implemented and performed using an analysis description language and its runtime interpreter CutLang, which allows quick analysis prototyping and histogramming~\cite{Unel:2019reo, Sekmen:2018ehb}.

The paper starts by introducing the down-type iso-singlet quark model in Section~\ref{sec:dmodel} followed by a description of the HL-LHC and FCC colliders and relevant experimental conditions in Section~\ref{sec:colliders}, and the analysis description language and runtime interpreter CutLang in Section~\ref{sec:cutlang}. Detailed explanation of the search for $D$ quarks and the search results are presented in Section~\ref{sec:analysis} followed by the conclusions in Section~\ref{sec:conclusions}.

\section{Down-type iso-singlet quark model}
\label{sec:dmodel}
If the group structure of the SM, $SU_C (3) \times SU_W (2) \times U_Y (1)$, originates from the breaking of the $E_6$ group at the GUT scale, then the quark sector of the SM is extended by the addition of an iso-singlet quark per family as:

\begin{eqnarray}
 \begin{pmatrix}  u_L\\ d_L \\
         \end{pmatrix} &,& u_R, d_R, D_L, D_R ;\\ \nonumber 
\begin{pmatrix}  c_L\\ s_L \\
         \end{pmatrix} &,& c_R, s_R, S_L, S_R ;\\ \nonumber
          \begin{pmatrix}  t_L\\ b_L \\
         \end{pmatrix} &,& t_R, b_R, B_L, B_R  .
\end{eqnarray}

In the considered model, the $S$ and $B$ quarks are assumed to be heavy and decoupled from the spectrum, leaving the $D$ quark as the only one accessible for searches at the present and near future colliders. A second assumption, following from the general behavior of CKM (Cabibbo, Kobayasi,  Maskawa), is that mixing inside a given family is stronger compared to mixing between different families.  Therefore, we only consider the Lagrangian relevant for the weak interaction of $d$ and $D$ quarks as given in \cite{_akir_1997}:

\begin{equation}
\begin{split}
L_D &= \frac{\sqrt{4\pi \alpha_{em}}}{2\sqrt{2}\sin{\theta_W}} \big[\bar{u}^{\theta} \gamma_\alpha (1 - \gamma_5) d \cos{\phi}\\
 &\quad + \bar{u}^{\theta} \gamma_\alpha (1 - \gamma_5) D \sin{\phi}\big] W^\alpha\\
 &\quad - \frac{\sqrt{4\pi \alpha_{em}}}{4\sin{\theta_W}}\left[\frac{\sin{\phi}\cos{\phi}}{\cos{\theta_W}} \bar{d}\gamma_\alpha (1 - \gamma_5) D\right] Z^\alpha\\
 &\quad - \frac{\sqrt{4\pi \alpha_{em}}}{4\cos{\theta_W}\sin{\theta_W}} \times \\
 &\qquad \quad \big[\bar{D} \gamma_\alpha (4 \sin^2{\theta_W} - 3 \sin^2{\phi}(1 - \gamma_5)) D\\
 &\qquad \quad + \bar{d}\gamma_\alpha(4 \sin^2{\theta_W} - 3 \sin^2{\phi}(1 - \gamma_5)) d\big] Z^\alpha \\
 & + h.c. \quad ,\label{lagrangian}
\end{split}
\end{equation}
where the superscript $\theta$ represents the usual CKM mixings taken to be in the up sector for simplicity of calculation, $\theta_W$ is the weak mixing angle and $\phi$ is the mixing angle between the $d$ and $D$ quarks, which is responsible for the decay of the $D$ quark. The limits on $\phi$ can be obtained from the current precision measurements for the $3\times 3$ CKM matrix elements, assuming that its $3\times 4$ extension has the sum of the squares of the elements of a row equal to 1. 

The evaluation of the presently measured values and their errors yield ${\lvert\sin{\phi}\rvert} \leq 0.035$ (0.043) allowing a 1(2) sigma variation on the first row elements~\cite{PhysRevD.98.030001}. The cross section calculation results for FCC are essentially insensitive to $\sin(\phi)$, since the studied pair production proceeds mostly via gluon exchange. However at HL-LHC, especially for large values of $D$ quark mass,  the production is mostly via the $q\bar{q}$ channel which has a slight $\sin(\phi)$ dependence for the cross section due to the $t$ channel sub-process propagating via $W$ boson as shown in \ref{fig:feynman}, sub-figure (d). Since this sub-process contributes with an opposite sign, reducing the mixing angle effectively increases the  $D\bar{D}$ production cross section.

The branching fractions for the three possible $D$ decay modes, ${D\rightarrow Wu}$, ${D\rightarrow Zd}$ and ${D\rightarrow hd}$ are about $50\%$, $25\%$ and $25\%$ respectively for masses above $\sim 800$~GeV~\cite{Sultansoy:2006cw}. In this study, we consider the pair production of $D$ quarks and their subsequent decay in the $D\rightarrow Zd$ channel to explore the discovery prospects of two possible future collider scenarios.

\section{Considered collider scenarios}
\label{sec:colliders}
\subsection{High-Luminosity LHC}
\label{sec:hllhc}

The LHC reached its design value of peak luminosity $10^{34}~cm^{-2} s^{-1}$ in June, 2016. The High-Luminosity Large Hadron Collider (HL-LHC) project aims to improve the performance of the LHC in order to increase the potential for discoveries after 2027~\cite{Apollinari:2120673, doi:10.1142/9581, ApollinariG.:2017ojx}. To implement this, HL-LHC will have several cutting-edge technologies, such as, 11–12 T superconducting  magnets; very compact  with  ultra-precise  phase  control superconducting  cavities  for  beam  rotation;  new  technology  for  beam  collimation;  and  long  high-power superconducting links with zero energy dissipation. HL-LHC is expected to reach the peak luminosity of $5 \times 10^{34}$ $cm^{-2} s^{-1}$, allowing an integrated luminosity of $250 fb^{-1}$ per year. Therefore, it gives an integrated luminosity of $3000~fb^{-1}$ in the operation period of about a dozen years after the upgrade. This integrated luminosity corresponds to ten times the amount LHC is expected to collect after 12 years of operation.

To meet the challenges brought by this higher luminosity at the HL-LHC, such as higher radiation dose, higher particle rate, higher pileup, and higher event rate, etc, the ATLAS and CMS detectors will undergo an extensive upgrade (i.e. the “Phase 2” upgrade). The ATLAS inner tracker (ITk) is being completely rebuilt for Phase 2, as a result of which, the pseudorapidity coverage will extend up to $|\eta| = 4$.  Moreover, new front-end electronics and a new readout system in the calorimeters will allow triggering higher resolution objects at the lowest trigger level at an increased rate, and lead to improved reconstruction. In addition, new inner barrel chambers will be installed in the muon detector system for increased coverage. The  CMS  detector  will  similarly  undergo  major  upgrades which include  a  replacement  of  the  silicon  strip  and  pixel components in  the  tracking detector increasing the coverage up to $|\eta|= 4$. The hadronic calorimeter will be read out by silicon photomultipliers.  The endcap electromagnetic and hadron calorimeters will be replaced with a new combined sampling calorimeter that will provide highly-segmented spatial information in both the transverse and longitudinal directions, as well as high-precision timing information. The  muon  system  will  be  extended  with  new  chambers  in  the  forward  region,  bringing  the  coverage  up  to $|\eta| = 2.8$.  Additionally, both ATLAS and CMS envisage adding timing detectors to provide the capability of adding timing information to reconstruction~\cite{CERN-LHCC-2015-020, Contardo:2020886}.

\subsection{Future Circular Collider}
\label{sec:fcc}

The Future Circular Collider (FCC) was launched as  a  world-wide international collaboration hosted at CERN in response to the 2013 Update of the European Strategy for Particle Physics (EPPSU)~\cite{Abada2019, Benedikt:2653674}. In the 2020 Update of EPPSU, it has been proposed to investigate the technical and financial feasibility of FCC~\cite{European:2720129}. FCC scenarios are studied for three different types of particle collisions, namely hadron (proton-proton and heavy ion), electron-positron and proton-electron collisions.  The proposed energy frontier proton-proton collider, FCC-hh, which is considered in this study, is designed to provide proton–proton collisions with a centre-of-mass energy of $100~TeV$ and an integrated luminosity of $20~ab^{-1}$ for 25 years of operation. The FCC-hh collider layout has two high luminosity interaction points for general purpose detectors. The factor 7 increase in energy over the present LHC requires a vast modification compared to the designs of current general purpose LHC detectors. The detectors for 100 TeV should be able to measure  multi-TeV jets, leptons and photons from heavy resonances with masses up to 50 TeV, while at the same time measuring the known SM processes with high precision, and still being sensitive to a broad range of BSM signatures with moderate momentum.  In addition, future detectors will need to operate at $~1000$ pileup events per bunch-crossing. The detector acceptance is targeted to increase up to $|\eta| = 4.4$ in order to improve sensitivity to vector boson fusion processes.

\section{CutLang analysis description language and runtime interpreter}
\label{sec:cutlang}

As mentioned earlier, one goal of this study is to test the feasibility of the new ``analysis description language" approach in analysis writing and running in phenomenological studies. An analysis description language is a domain-specific, declarative language designed to express the physics contents of an analysis in a standard and unambiguous way. In this approach, the description of the analysis components is decoupled from the software framework that run the analysis. 

This study uses the language ADL~\cite{Brooijmans:2016vro, Brooijmans:2018xbu, Brooijmans:2020yij}, which consists of a plain text file containing blocks with a keyword-value structure. The blocks make clear the separation of analysis components such as object definitions, variable definitions, and event selections while the keywords specify analysis concepts and operations. The syntax includes mathematical and logical operations, comparison and optimization operators, reducers, four-vector algebra and common HEP-specific functions (e.g. $\delta\phi$, $\delta R$, etc.). ADL files can refer to self-contained functions encapsulating variables
with complex algorithms (e.g. $M_{T2}$, aplanarity, etc.) or non-analytic variables (e.g. efficiency tables, machine learning discriminators, etc.). 

ADL can be used for performing an analysis by any framework capable of interpreting and running it.  Here, we use CutLang~\cite{Sekmen:2018ehb, Unel:2019reo}, a runtime interpreter, which is able to operate directly on events without the need for compilation. CutLang is written in C++ and is based on ROOT~\cite{ROOT} classes for Lorentz vector operations and histogramming.  
It uses automatically generated dictionaries and grammar rules based on unix tools Lex and Yacc~\cite{lexandyacc} . The typical output of an analysis in CutLang is a file containing surviving events and histograms which can be used for statistical analysis. 
 
Not having the necessity to write or compile code, combined with the simple, human-readable nature of ADL syntax makes it a very practical construct for quickly performing phenomenological analyses such as the one in this study.  

\section{Search for down-type iso-singlet quarks}
\label{sec:analysis}

\subsection{Signal and background processes}

\begin{figure}[h!]
  \begin{subfigure}[H]{0.49\linewidth}
    \centering
      \includegraphics{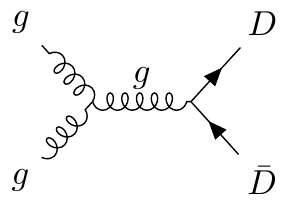}
    \caption{gluons, s channel}
  \end{subfigure}
  \begin{subfigure}[H]{0.49\linewidth}
    \centering
     \includegraphics{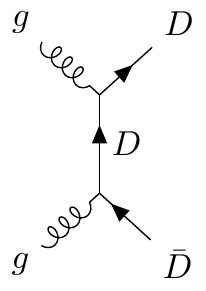}
    \caption{gluons, t channel}
  \end{subfigure}
  \newline
  \begin{subfigure}[H]{0.49\linewidth}
   \centering
   \includegraphics{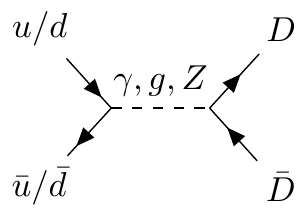}
    \caption{up quarks, s channel}
  \end{subfigure}
  \begin{subfigure}[H]{0.49\linewidth}
   \centering
    \includegraphics{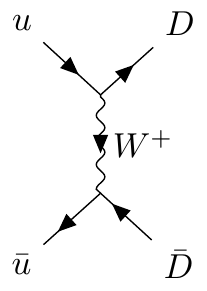}
    \caption{up quarks, t channel}
  \end{subfigure}
  \caption{Tree level Feynman Diagrams for the process $pp\rightarrow D\bar{D}$ }
  \label{fig:feynman}   
\end{figure}

\begin{figure}[h!]
\centering
  \includegraphics[width=\linewidth]{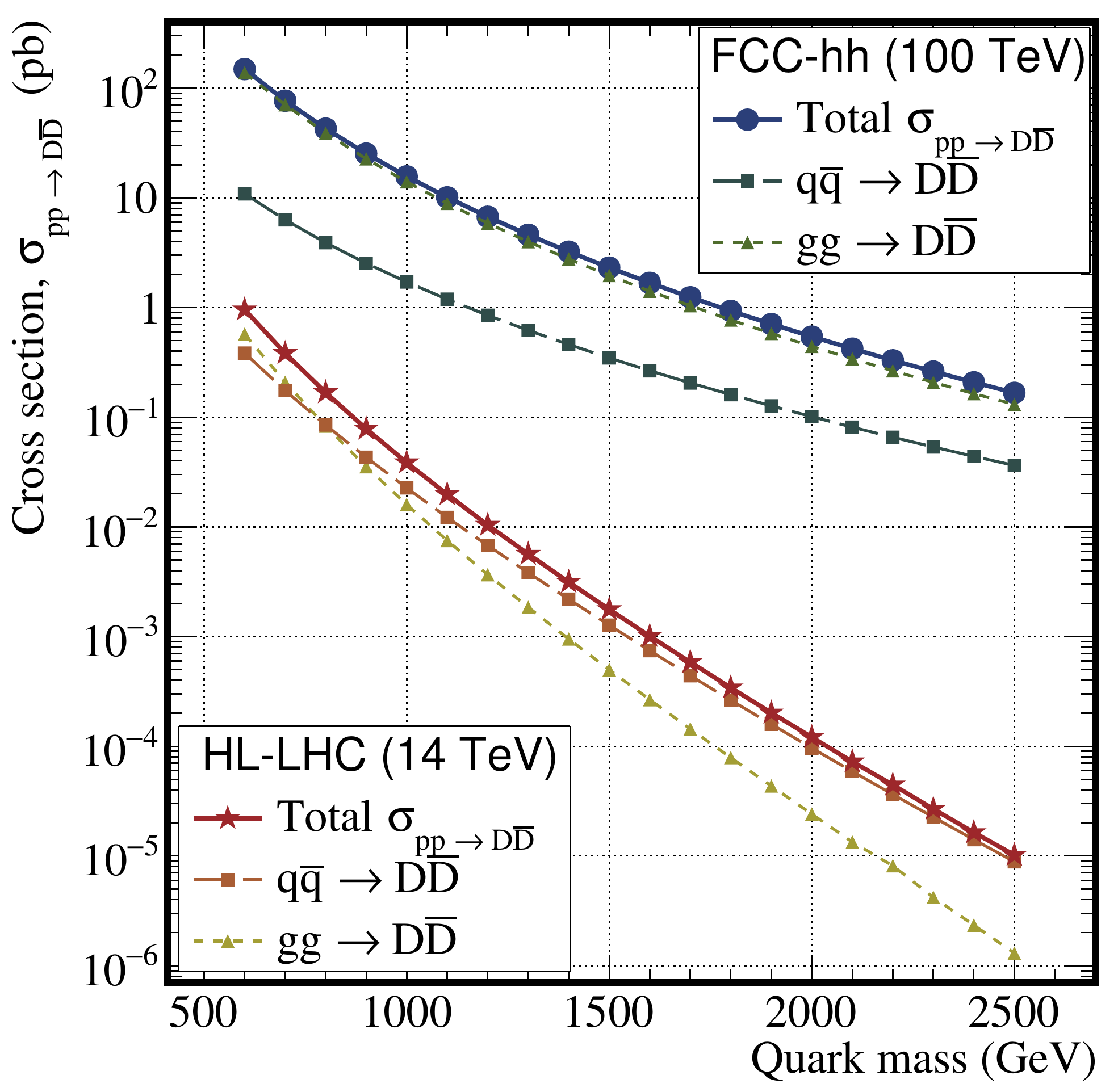}
\caption{$pp \rightarrow D\bar{D}$, $q\bar{q} \rightarrow D\bar{D}$ and $gg \rightarrow D\bar{D}$ cross sections vs $D$ quark mass for HL-LHC and FCC-hh energies, calculated using CompHep. The $d-D$ mixing angle is taken as $\sin{\phi} = 0.035$}
\label{fig:cross-section}       
\end{figure}

The main tree level Feynman diagrams for the pair production of $D$ quarks at hadron colliders are presented in Figure~\ref{fig:feynman}. The model Lagrangian in equation \eqref{lagrangian} was implemented into the tree level event generator, \textit{CompHEP} \cite{BOOS2004250, Pukhov:1999gg}. The resulting pair production cross sections at generator level for HL-LHC and FCC-hh for the $gg$ and $q\bar{q}$ channels and their sum are shown in Figure~\ref{fig:cross-section} as a function of $D$ quark mass. {  The pair production cross section is somewhat smaller than the single production, for example for a $D$ quark of 1 TeV the former is 38.6 fb whereas it is 94.5 fb for the latter. However as the single production results depend heavily on the mixing angle and the SM background is especially large due to QCD jets this paper focuses on pair production. }


The $E_6$ GUT model does not predict the masses of the iso-singlet quarks. Therefore, this study scans some plausible values for the $D$ quark mass (up to $2500$~GeV) to estimate the experimental reach at both HL-LHC and FCC-hh machines.  The iso-singlet quarks are expected to immediately decay into SM particles due to their large masses. In this analysis, we have considered the decay process  $D\bar{D}\rightarrow ZZd\bar{d}$, with subsequent leptonic decays of both $Z$ bosons, $Z \rightarrow \ell^+\ell^-$. 

The main SM  background to the signal process is  $pp \rightarrow ZZjj$ production, with subsequent leptonic decays of both Z bosons. The SM cross-section of $pp \rightarrow ZZjj$ is calculated using \texttt{MadGraph5\_aMC@NLO}~\cite{Alwall2014}  considering up to 4 QED and QCD interaction vertices and found to be $2.918$~pb and $68.04$~pb for HL-LHC and FCC-hh, respectively.

Processes with Higgs decaying to two $Z$ bosons also provide final states resembling that of the signal, however they are not considered as significant backgrounds in this study due to relatively low effective cross sections as well as one of the $Z$ bosons from the Higgs boson decay being virtual.  At 14 TeV, the Higgs production cross sections are estimated as 54.6~pb from gluon fusion, 4.3~pb from VBF, 1.5~pb from $WH$, 0.98~pb from $ZH$ and 0.55~pb from $bbH$ production channels. 
To obtain an estimate for $ZZjj$ final states, these numbers are multiplied by the $h\rightarrow ZZ$ branching fraction and the hadronic branching fraction of $W$ and $Z$ bosons. Moreover, the gluon fusion cross section is corrected to account for multi-jet events~\cite{Greiner:2124367}. Extrapolating linearly from 8 and 13 TeV results, the $h+2j$ cross section from gluon fusion is estimated as 5.4~pb at NLO level. Folding in the appropriate branching fractions, the total effective cross section for the Higgs-related backgrounds becomes $\sim$0.31~pb, which is a small fraction of the direct $ZZjj$ production cross section.  The approximate estimate of these processes for 100 TeV is $\sim$5~pb, which is similarly small compared to the SM $ZZjj$ cross section.  Moreover, one of the $Z$ bosons originating from the Higgs decays would be virtual.  Therefore the majority of such events would be rejected by the requirement of two reconstructed $Z$ bosons having an invariant mass of 91.2 GeV in our analysis.

The $E_6$ model signal events with $D$ quarks decaying to SM particles and SM background events were generated using  \textit{CompHEP} and \textit{MadGraph5\_aMC@NLO} respectively. The CompHEP setup was adjusted to impose a generator level requirement of 10~GeV on the transverse momenta of the SM d-quarks originating from the $D\rightarrow Zd$. The NNPDF 3.1 parton distribution function set~\cite{Ball:2017nwa}, which is the most up-to-date set available has been used both for 14 and 100 TeV. Further decays and showering and hadronization processes were simulated using \textit{Pythia6}~\cite{Sjostrand:2006za}. Pythia was set up to only allow electron and muon decays of the $Z$ bosons. Subsequently, the detector effects were modelled with the fast detector simulation program \textit{Delphes}~\cite{deFavereau2014} using the configurations~\cite{RefHL-LHCcard} and~\cite{RefFCCcard} for generic HL-LHC and FCC-hh detectors.  

\subsection{Object and event reconstruction and selection}

The complete object and event reconstruction and selection algorithm for the analysis is given in ADL format in Table~\ref{tab:CutLang}. This is, in fact, the exact ADL code run in CutLang to produce the results presented in this paper.  

The analysis is performed in the $4\ell+2j$ channel, and thus uses leptons and jets. Both for HL-LHC and FCC-hh cases, leptons considered are electrons and muons, which are both required to have transverse momentum $p_T > 20$~GeV and pseudorapidity $|\eta| < 4$. { Electrons (muons) are required to have an isolation of 0.1 (0.2) within a cone of $dR < 0.3$. }
Jets are reconstructed with the anti-$k_T$ algorithm with a radius of $R=0.5$, and are required to have $p_T > 50$~GeV (which is higher than generator level requirement) and $|\eta| < 4$.  Increased pseudorapidity acceptance at the HL-LHC and FCC-hh detectors compared to LHC will provide an increased sensitivity for the analysis. Events are required to have at least 4 leptons and at least 2 jets as defined above.

\subsubsection{Leptonic Z boson reconstruction}
The two $Z$ boson candidates from the $D$ decay are reconstructed from the selected leptons. For an efficient Z boson reconstruction, we consider the following criteria:

\begin{center}
\begin{enumerate}
    \item Mass of the reconstructed Z boson candidate should be as close as possible to 91.2 GeV,
    \vspace{2mm}
    \item the Z boson candidate should be flavour and charge neutral (i.e, reconstructed from a $e^+e^-$ or a $\mu^+ \mu^-$ pair)
\end{enumerate}
\end{center}

{ In this analysis, we are focused on final states with Z bosons with moderate momentum, which decay to non-collimated leptons that can be independently reconstructed.  However, especially at the FCC-hh energies, higher mass $D$ quarks  yield a Z boson $p_T$ spectra with a higher component of boosted Z bosons that would decay to collimated lepton pairs.  Such collimated lepton pairs would partially fail to be identified as two individual leptons due to the lepton isolation requirement and be counted as a single lepton, resulting in the event failing the 4 lepton criteria.  A more effective treatment of the boosted final states would require Z boson reconstruction via explicit tagging of the boosted Z boson via collimated lepton jets.  These boosted channels can be added when collimated lepton jet tagging performance or simulation for the FCC-hh conditions become available, and they would increase the analysis sensitivity.}

\begin{figure}[!h]
\centering
  \includegraphics[width=.9\linewidth]{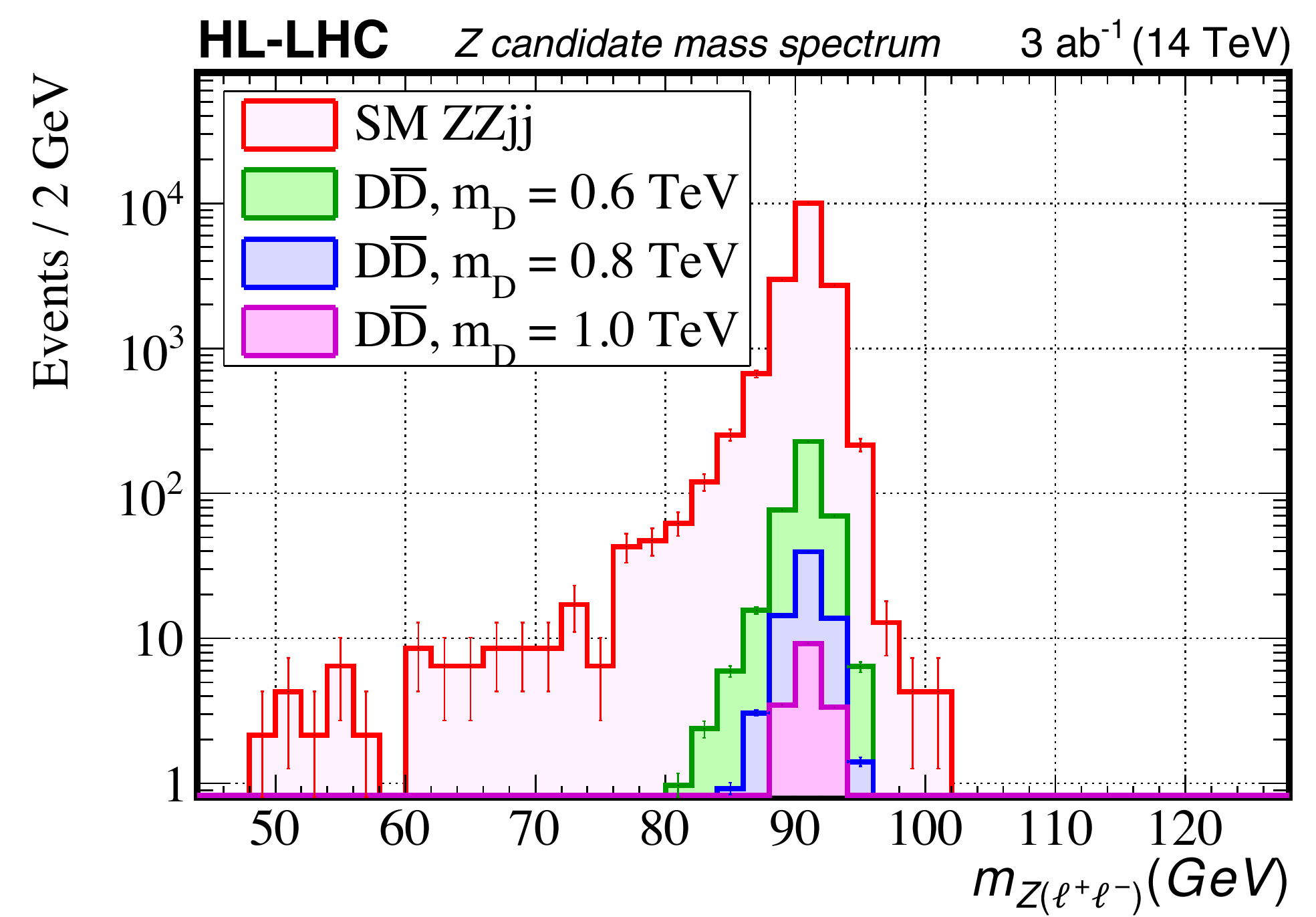}
  \includegraphics[width=.9\linewidth]{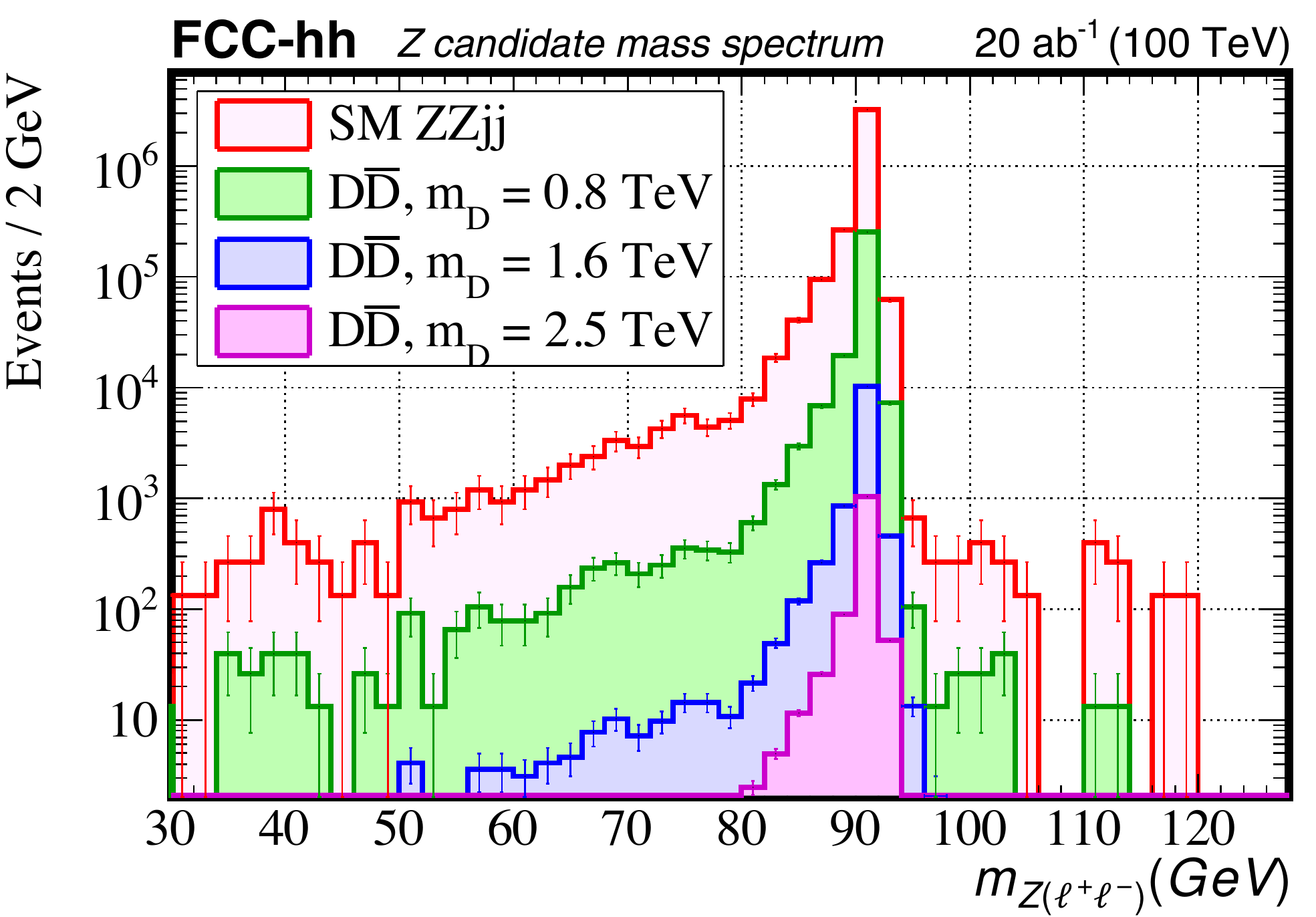}
\caption{Invariant mass distribution for both reconstructed $Z$ boson candidates for HL-LHC (top) and FCC-hh (bottom) conditions.  Candidates are reconstructed from both $e^+e^-$ and $\mu^+\mu^-$ pairs.}
\label{fig:mZ}
\end{figure}

For the resolved final state, leptons are paired to reconstruct both $Z$ bosons simultaneously in the $\chi^2$ expression below, which both selects the dilepton combinations with masses as close as possible to the measured $Z$ mass of 91.2 GeV and ensures the same flavor requirement on dileptons in a candidate:  
\begin{eqnarray}\label{chidef}
\nonumber \chi^2_{ZZ} &\equiv& (m_{Z1} - 91.2)^2 + (m_{Z2} - 91.2)^2\\ \nonumber
&+& \left(999\times PdgID\left[Z_1\right]\right)^{2} + \left(999 \times PdgID\left[Z_2\right]\right)^{2}.\\
\end{eqnarray}
More information on technical implementation of Z reconstruction and the $\chi^2_{ZZ}$ in ADL and CutLang is given in Appendix~\ref{sec:appendix}.  The reconstructed $Z$ candidates are additionally required to have a total electric charge of 0.  Mass distributions of both $Z$ candidates reconstructed from $e^+e^-$ and $\mu^+\mu^-$ pairs are shown in Figure~\ref{fig:mZ} for different $D$ quark masses for HL-LHC and FCC-hh.

\subsubsection{$D$ quark reconstruction}

\begin{figure}[!h]
\centering
  \includegraphics[width=.9\linewidth]{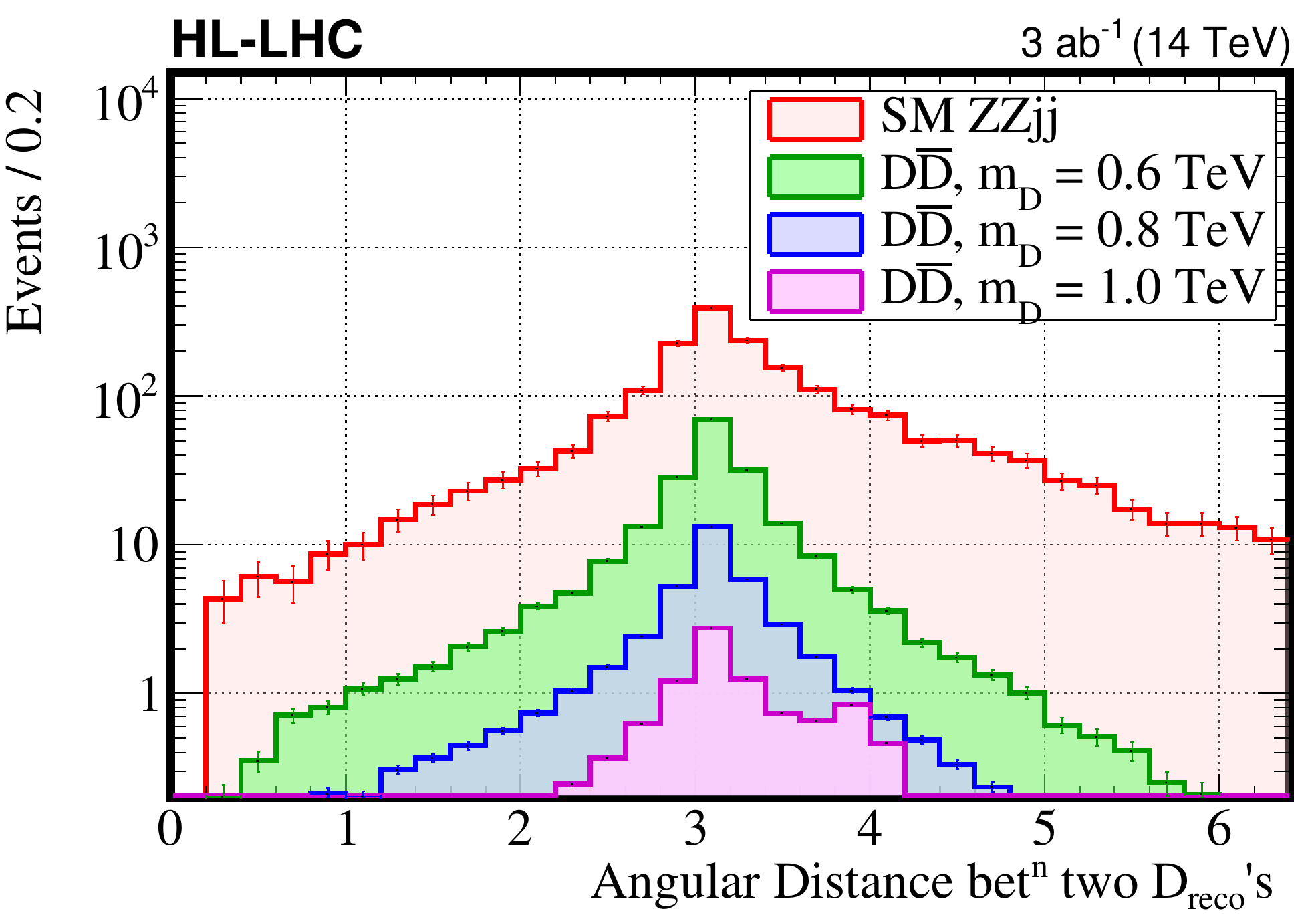}
  \includegraphics[width=.9\linewidth]{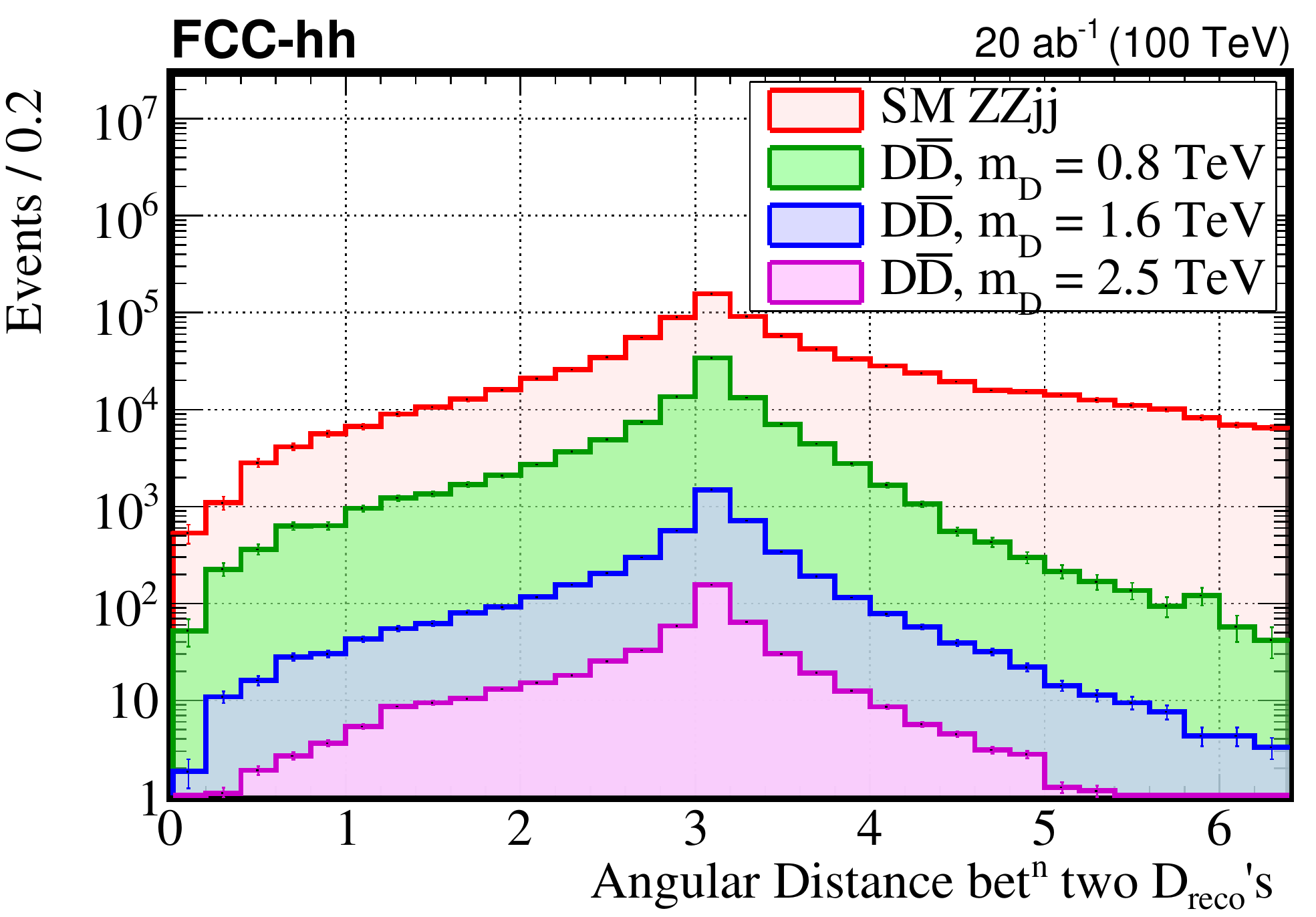}
\caption{Distribution of angular distance between the two reconstructed $D$ quark candidates $D_1$ and $D_2$ for HL-LHC (top) and FCC-hh (bottom) conditions. The jets are selected by minimizing the condition defined in Eq.~\ref{eq:chi2mD}.}
\label{fig:dRDD}
\end{figure}

Each $D$ quark candidate ($D_1$ and $D_2$) is reconstructed from a $Z$ boson candidate and a jet.
Once again, the reconstruction is based on a $\chi^2$ optimization which takes into account the following conditions:
\vspace{1mm}
\begin{center}
\begin{enumerate}
\item $D$ quark mass is presumed unknown. However, masses of the two reconstructed $D$ quark candidates should be as close as possible to each other. We express this condition as:
\begin{equation}
 \chi^2_{m_D} \equiv ((m_{D_1} - m_{D_2})/m_D )^2 \; ,
 \label{eq:chi2mD} 
\end{equation}
where $m_{D_1}$ and $m_{D_2}$ are the invariant masses of the two $D$ quark candidates and $m_D = (m_{D_1} + m_{D_2})/2$.
\vspace{1mm}
\item Transverse momentum of the jets directly originating from the $D$ quark decay is expected to be high. To ensure selecting jets with high momentum, we use the Heavyside step function with a weight factor:
\begin{eqnarray}
\nonumber \chi^2_{p_{T,j}} & \equiv & H(p_{T,j}^{cut} - p_{T,j_1}) \times ((p_{T,j}^{cut} / p_{T,j_1}) - 1.0) \\
 & + & H(p_{T,j}^{cut} - p_{T,j_2}) \times ((p_{T,j}^{cut} / p_{T,j_2}) - 1.0),
  \label{eq:chi2pTj}
\end{eqnarray}
where $p_{T,j_1}$ and $p_{T,j_2}$ are the transverse momenta of the jets and $p_{T,j}^{cut}$ is the selection threshold to be applied to the jet transverse momenta.  To determine the optimal value for this threshold which would obtain the best signal-background separation, we show the $p_T$ distributions of the candidate jets in Figure~\ref{fig:ptj} for signals with different $m_D$ and the background at HL-LHC and FCC-hh.  The jets are selected by minimizing the condition defined in Eq.~\ref{eq:chi2mD}.  Due to its much higher center-of-mass energy, FCC-hh yields a much harder jet $p_T$ spectrum. Based on these distributions, we select $p_{T,j}^{cut}= 300$ and $500$~GeV as thresholds for HL-LHC and FCC-hh respectively.
\begin{figure}[!h]
\centering
  \includegraphics[width=.9\linewidth]{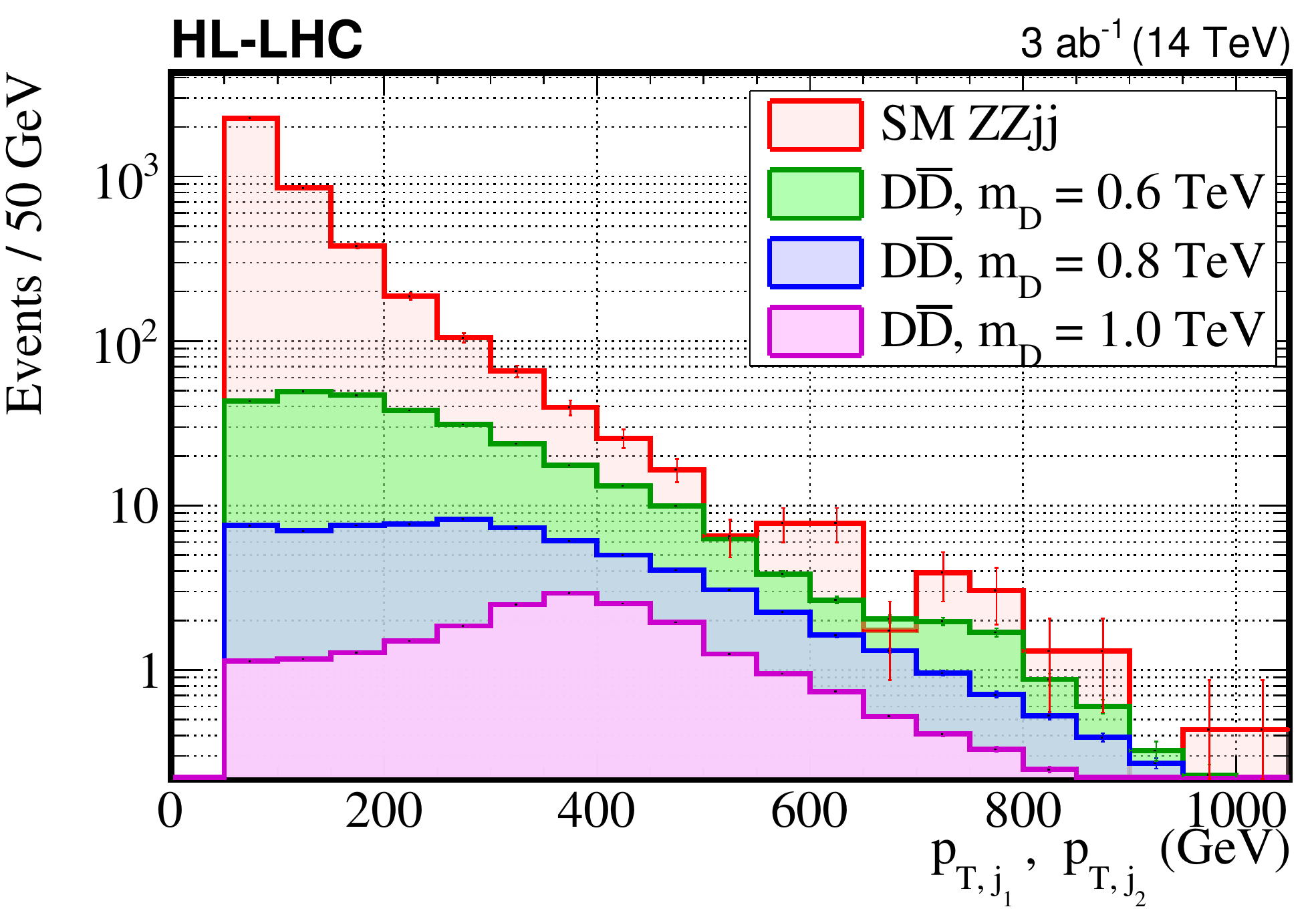}
  \includegraphics[width=.9\linewidth]{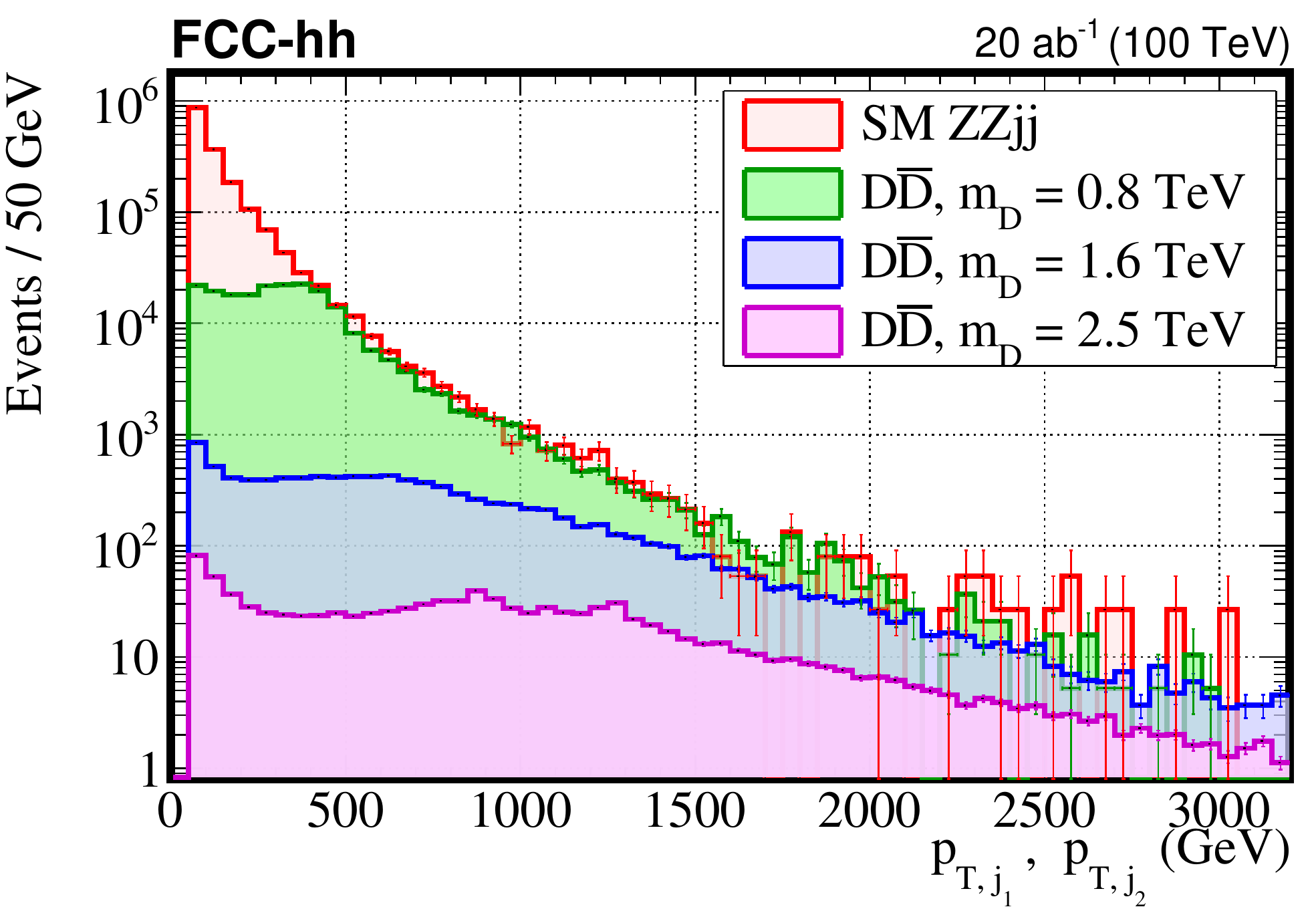}
\caption{ Transverse momentum distribution for both jets used in $D$ quark reconstruction for HL-LHC (top) and FCC-hh (bottom) conditions. The jets are selected by minimizing the condition defined in Eq.~\ref{eq:chi2mD}.}
\label{fig:ptj}
\end{figure}
\vspace{1mm}
\item Angular separation between the two $D$ quarks,
\begin{equation}
dR_{DD} = \sqrt{(\eta_{D_1} - \eta_{D_2})^2 + (\phi_{D_1} - \phi_{D_2})^2} \; ,
\end{equation}
should reflect that the $D$ quarks are centrally produced, with negligible Lorentz boost.  The most characteristic configuration would correspond to $D$ quarks having $|\eta| \simeq 0$ and being back-to-back on the transverse plane, which gives $\delta\phi \simeq \pi$, where $\delta\phi$ represents the $\phi$ difference of the two particles.  As a result, $dR$ is expected to be dominated by $\delta\phi$ and peak around 3.14.  This can be seen in Figure~\ref{fig:dRDD}, which shows the $dR$ distributions for signals and the background for HL-LHC and FCC-hh, after applying a minimization based on Eq~\ref{eq:chi2mD}.  Both signals and the background peak around 3.14, but the backgrounds display a wider distribution. Based on this information, we define a variable that can be minimized to zero: 
\begin{eqnarray}
\chi^2_{dR_{DD}} & \equiv & (dR_{DD}/3.14 - 1.0)^2.
 \label{eq:chi2dRDD}
\end{eqnarray}

\end{enumerate}
\end{center}

We then combine the three conditions in Eqs~\ref{eq:chi2mD},~\ref{eq:chi2pTj} and ~\ref{eq:chi2dRDD} to obtain a $\chi^2$ and select the $D$ candidates by running a minimization based on the sum:
\begin{equation}\label{eq:chiDD}
    \chi^2_{DD} \equiv \chi^2_{m_D} + \chi^2_{p_{T,j}} + \chi^2_{dR_{DD}} \simeq 0 \quad.
\end{equation}

{Here, we tried different relative weighting of $\chi^2_{m_D}$, $\chi^2_{p_{T,j}}$ and $\chi^2_{dR_{DD}}$, but the above choice gives the optimal result.}  
\subsubsection{Final selection on $\chi_{DD}^2$ }

\begin{figure}[ht!]
  \includegraphics[width=.9\linewidth]{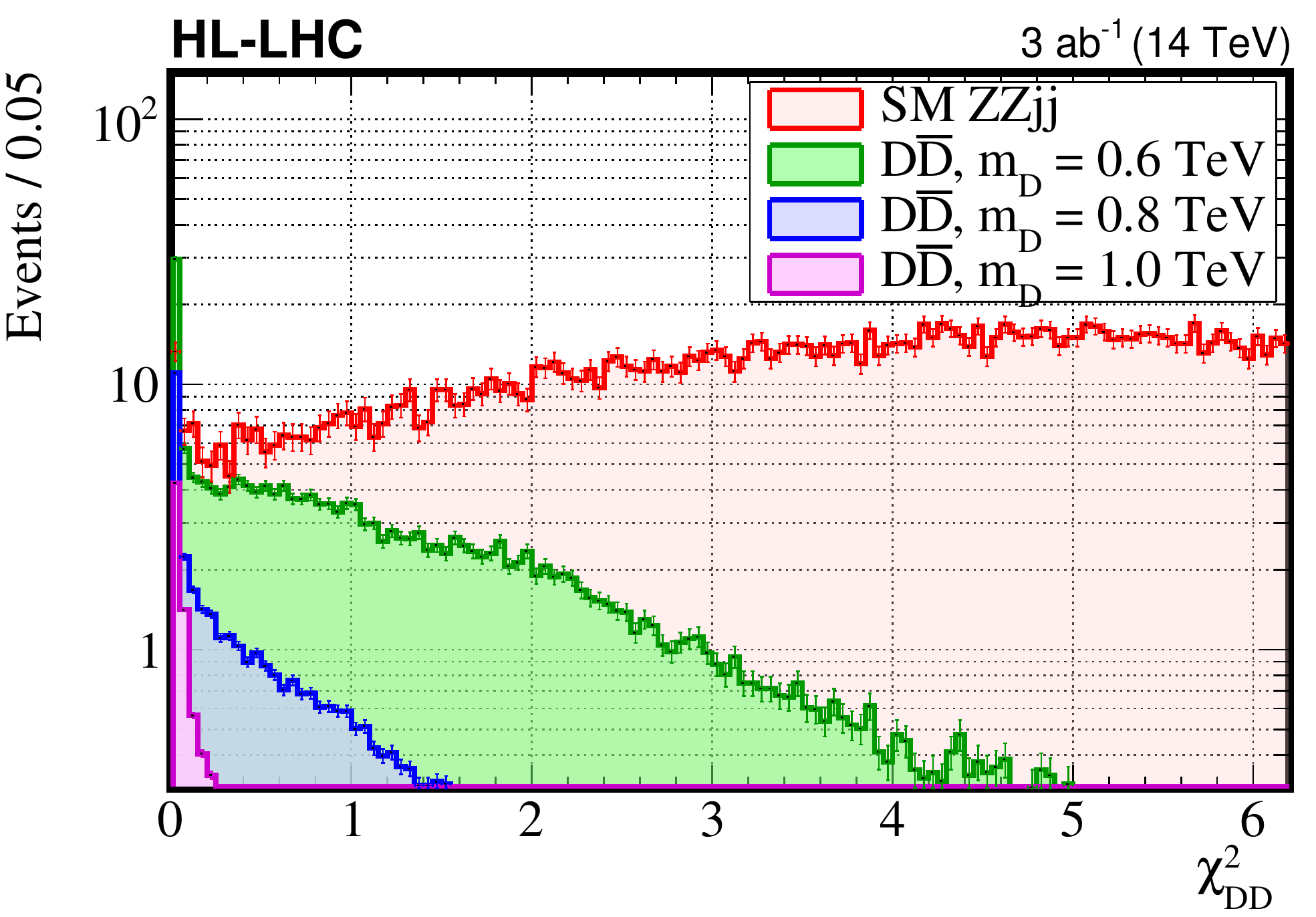}
  \includegraphics[width=.9\linewidth]{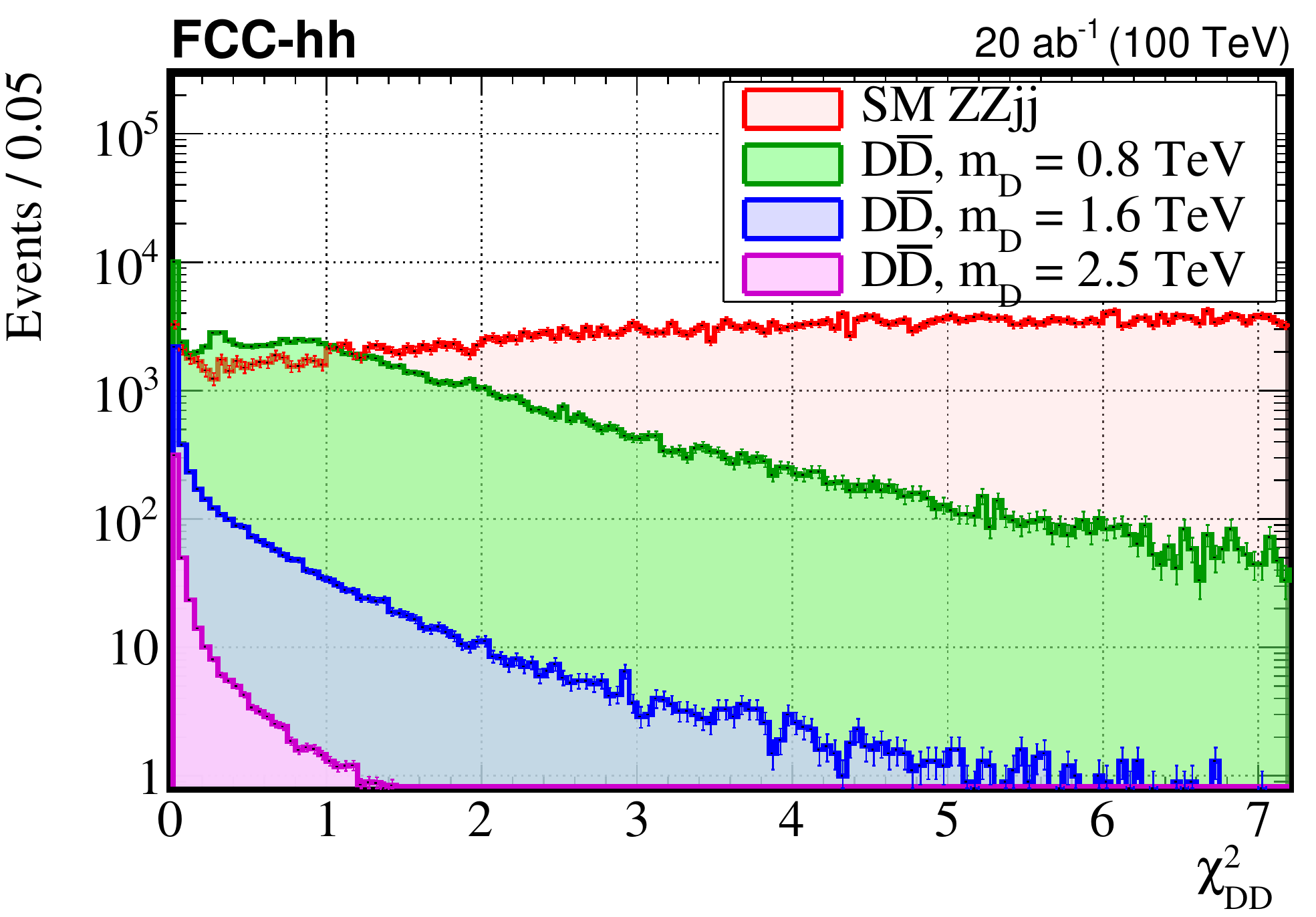}
\caption{Distribution of $\chi^2_{DD}$ values obtained after minimization for HL-LHC (top) and FCC-hh (bottom) conditions.} 
\label{fig:chi2DD}
\end{figure}

Figure~\ref{fig:chi2DD} shows the distribution of $\chi^2_{DD}$ values obtained after minimization for HL-LHC (top) and FCC-hh (bottom) conditions for signals with different $m_D$ and background.  As expected, the signals exhibit a distribution much closer to zero compared to the background. A selection of $\chi_{DD}^2 < 0.5$ was applied to further reduce the SM contamination. The threshold value was chosen to ensure a high signal significance.

\begin{table*}[htbp]
\caption{Analysis description using the ADL/CutLang syntax.  This description can be directly processed with CutLang over events.}
\label{tab:CutLang}
\begin{lstlisting}[breaklines, frame=single]
## Object definitions
object goodJet
  take JET
  <@\textcolor{Fuchsia}{select}@> Pt(JET) > 50 
  <@\textcolor{Fuchsia}{select}@> <@\textcolor{black}{abs}@>(Eta(JET) < 4 
object goodEle
  take ELE
  <@\textcolor{Fuchsia}{select}@> Pt(ELE) > 20
  <@\textcolor{Fuchsia}{select}@> <@\textcolor{black}{abs}@>(Eta(ELE)) < 4
object goodMuo 
  take MUO
  <@\textcolor{Fuchsia}{select}@> Pt(MUO) > 20
  <@\textcolor{Fuchsia}{select}@> <@\textcolor{black}{abs}@>(Eta(MUO)) < 4 
object goodLep : Union (goodEle,goodMuo)
## Reconstructed particles
<@\textcolor{violet}{define}@> Zreco1 = goodLep[-1] goodLep[-1] <@\label{line:ZrecoDef}@>
<@\textcolor{violet}{define}@> Zreco2 = goodLep[-3] goodLep[-3]
<@\textcolor{violet}{define}@> dj1     <@\hspace{0.05cm}@>= goodJet[-2]
<@\textcolor{violet}{define}@> dj2     <@\hspace{0.05cm}@>= goodJet[-4]
<@\textcolor{violet}{define}@> Dreco1 = Zreco1 dj1 <@\label{line:DrecoDef}@>
<@\textcolor{violet}{define}@> Dreco2 = Zreco2 dj2
## Event variables
<@\textcolor{violet}{define}@> PTj1   = Pt(dj1)
<@\textcolor{violet}{define}@> PTj2   = Pt(dj2)
<@\textcolor{violet}{define}@> mZ1    = m(Zreco1)
<@\textcolor{violet}{define}@> mZ2    = m(Zreco2)
<@\textcolor{violet}{define}@> mD1    = m(Dreco1)
<@\textcolor{violet}{define}@> mD2    = m(Dreco2)
<@\textcolor{violet}{define}@> mD     = ( mD1 + mD2 ) / 2
<@\textcolor{violet}{define}@> dRDD   = dR( Dreco1 , Dreco2 )
# <@\textcolor{twilightlavender}{define}@> <@\textcolor{Gray}{PTjcut      = 300}@>  # for HL-LHC 
<@\textcolor{violet}{define}@> PTjcut      = 500 # for FCC-hh 
<@\textcolor{violet}{define}@> chi2DDcut = 0.5
## Chi2 variable definitions <@\hspace{0.2cm}@> 
<@\textcolor{violet}{define}@> chi2ZZ <@\hspace{0.18cm}@>= (mZ1 - 91.2)^2 + (mZ2 - 91.2)^2 + (999*pdgID(Zreco1))^2 + (999*pdgID(Zreco2))^2
<@\textcolor{violet}{define}@> chi2mD <@\hspace{0.15cm}@>= ((mD1 - mD2)/mD)^2
<@\textcolor{violet}{define}@> chi2PTj = Hstep(PTjcut -PTj1)*(PTjcut/PTj1 - 1.0) + Hstep(PTjcut - PTj2)*(PTjcut/PTj2 - 1.0) 
<@\textcolor{violet}{define}@> chidRDD = (dRDD/3.14 - 1.0)^2
<@\textcolor{violet}{define}@> chi2DD <@\hspace{0.15cm}@>= chimD + chiPTj + chidRDD 
## Event selection
<@\textcolor{RoyalBlue}{region}@> <@\textcolor{black}{DDselection}@>
  <@\textcolor{Fuchsia}{select}@>  ALL
  <@\textcolor{Fuchsia}{select}@>  Size(goodEle) >= 0
  <@\textcolor{Fuchsia}{select}@>  Size(goodMuo) >= 0
  <@\textcolor{Fuchsia}{select}@>  Size(goodLep) >=  4<@\label{line:LeptonSelect}@>
  <@\textcolor{Fuchsia}{select}@>  chi2ZZ ~= 0<@\label{line:ZrecoSearch}@>
  ## chi^2 optimization for Z reconstruction 
  <@\textcolor{Fuchsia}{select}@>  q(Zreco1) == 0         # Z is neutral
  <@\textcolor{Fuchsia}{select}@>  q(Zreco2) == 0         # Z is neutral
  <@\textcolor{Fuchsia}{histo}@>   hmZ1,  "Z candidate1 mass (GeV)", 320, 0.0, 3200.0,  mZ1
  <@\textcolor{Fuchsia}{histo}@>   hmZ2,  "Z candidate2 mass (GeV)", 320, 0.0, 3200.0,  mZ2      # mZ1 & mZ2 histogram plotting (fig. 3)
  <@\textcolor{Fuchsia}{select}@>  Size ( goodJet ) >= 2
  <@\textcolor{Fuchsia}{select}@>  chi2DD ~= 0<@\label{line:DrecoSearch}@>
  ## chi^2 optimization for D reconstruction <@\hspace{0.2cm}@>
  <@\textcolor{Fuchsia}{histo}@>   hchi2DD,   "chi2DD ", 200, 0.0, 10.0,  chi2DD                 # chi2DD histogram plotting (fig. 6)
  <@\textcolor{Fuchsia}{select}@>  chi2DD < chi2DDcut 
  <@\textcolor{Fuchsia}{histo}@>   hmD,  "D candidate mass (GeV)", 320, 0.0, 3200.0,  mD         # mD histogram plotting (fig. 7 & 8)
\end{lstlisting}
\end{table*}

\begin{table*}[htbp]
\centering
\caption{Percentage selection efficiencies for various signals and background for the HL-LHC selection.}
\label{tab:efficiencyHLLHC}       
\begin{tabular}{lcccc}
\hline
\multicolumn{1}{c}{\multirow{3}{*}{\begin{tabular}[c]{@{}c@{}}Cumulative \\ Selection criteria\end{tabular}}} & \multicolumn{4}{c}{\begin{tabular}[c]{@{}c@{}} Selected events (\% of Total)\end{tabular}} \\ \cline{2-5} 
\multicolumn{1}{c}{}                                                                                          & \multirow{2}{*}{Background}             & \multicolumn{3}{c}{Signal}                                       \\ \cline{3-5} 
\multicolumn{1}{c}{}                                                                                          &                                         & 600 GeV             & 800 GeV             & 1000 GeV             \\ \hline
\verb!ALL!               & 100      & 100     & 100      & 100                 \\
\verb!Size(goodLep) >= 4! & 20.1     & 26.7      & 28.1    & 30.0   \\
\verb!chi2ZZ ~= 0!         & 20.1     &26.7   &28.1    &30.0   \\
\verb!{Zreco1}q == 0!    & 20.1     & 26.7  & 28.1  & 30.0 \\
\verb!{Zreco2}q == 0!    & 20.1    & 26.7  & 28.1  & 30.0   \\
\verb!Size(goodJet) >= 2!&5.03     & 25.0  & 26.9   & 29.3  \\
\verb!chi2DD ~= 0!         &5.03     & 25.0    & 26.9     & 29.3    \\
\verb!chi2DD < chi2DDcut!     & 0.187      & 8.18  & 15.2    & 22.5                  \\
\hline
\end{tabular}
\end{table*}

\begin{table*}[htbp]
\centering
\caption{Percentage selection efficiencies for various signals and background for the FCC-hh selection.}
\label{tab:efficiencyFCC}       
\begin{tabular}{lcccc}
\hline
\multicolumn{1}{c}{\multirow{3}{*}{\begin{tabular}[c]{@{}c@{}}Cumulative \\ Selection criteria\end{tabular}}} & \multicolumn{4}{c}{\begin{tabular}[c]{@{}c@{}} Selected events (\% of Total)\end{tabular}} \\ \cline{2-5} 
\multicolumn{1}{c}{}                                                                                          & \multirow{2}{*}{Background}             & \multicolumn{3}{c}{Signal}                                       \\ \cline{3-5} 
\multicolumn{1}{c}{}                                                                                          &                                         & 800 GeV             & 1600 GeV             & 2500 GeV             \\ \hline
\verb!ALL!     & 100    & 100   & 100     & 100                 \\
\verb!Size(goodLep) >= 4! & 28.3      & 42.2    & 48.6    & 51.6     \\
\verb!chi2ZZ ~= 0!         & 28.3     &42.2    &48.6    &51.6    \\
\verb!{Zreco1}q == 0!    & 28.3      & 42.2   & 48.6    & 51.6    \\
 \verb!{Zreco2}q == 0!   & 28.3       & 42.2  & 48.6    & 51.6     \\
\verb!Size ( goodJet ) >= 2!   &13.2  & 41.0  & 48.1    & 51.3      \\
\verb!chi2DD ~= 0!         &13.2    & 41.0    & 48.1     & 51.3    \\
\verb!chi2DD < chi2DDcut!    & 0.302    & 11.2     & 36.2      & 46.0 \\
\hline
\end{tabular}
\end{table*}
\subsection{Results}

The percentage selection efficiencies for signal and background events for the event selection criteria described above are given in Tables~\ref{tab:efficiencyHLLHC} and~\ref{tab:efficiencyFCC} for HL-LHC and FCC-hh. Overall signal selection efficiency is seen to increase as $D$ mass increases.

\begin{figure}[!h]
\centering
  \includegraphics[width=.9\linewidth]{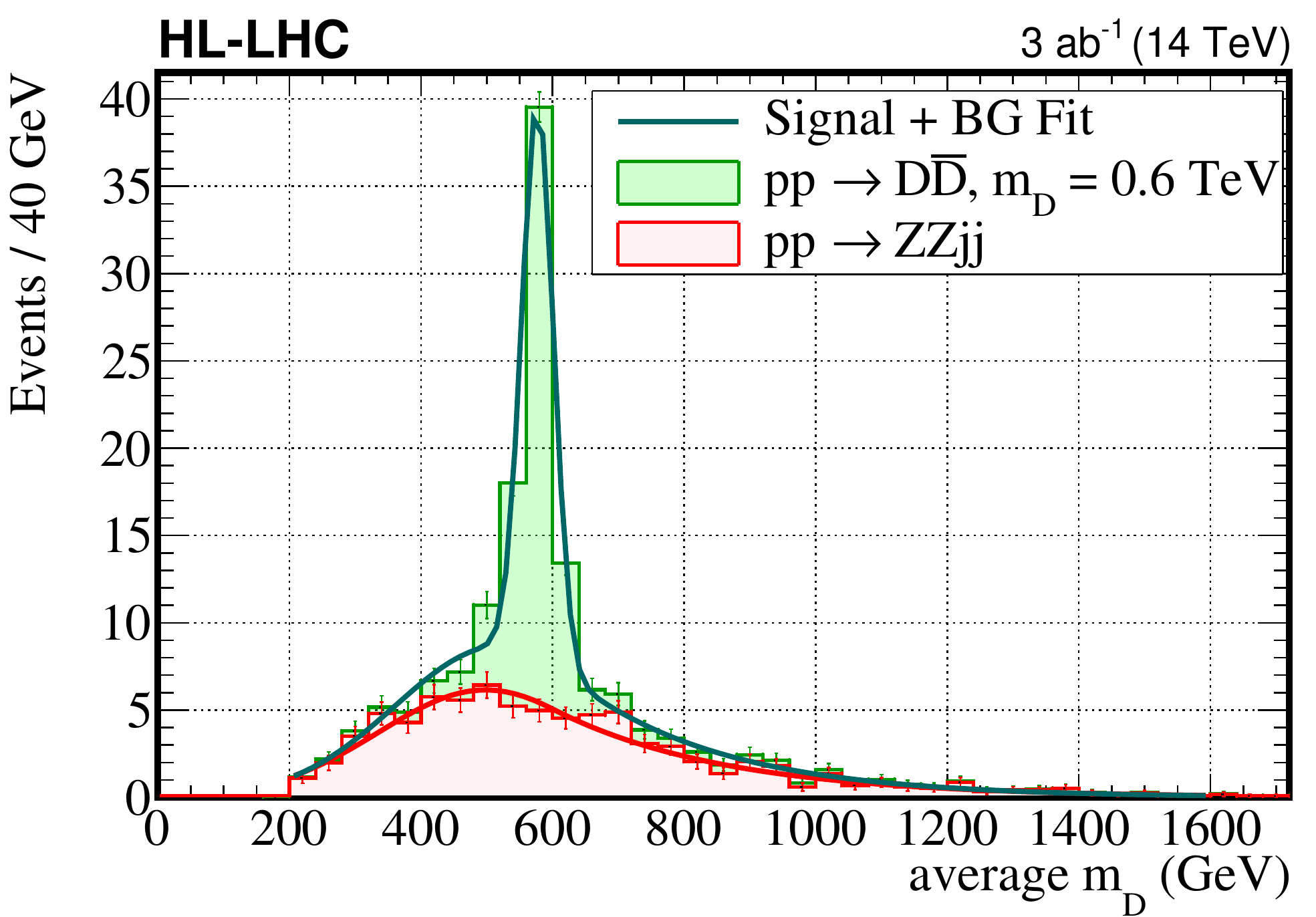}
\label{fig:4}       
  \includegraphics[width=.9\linewidth]{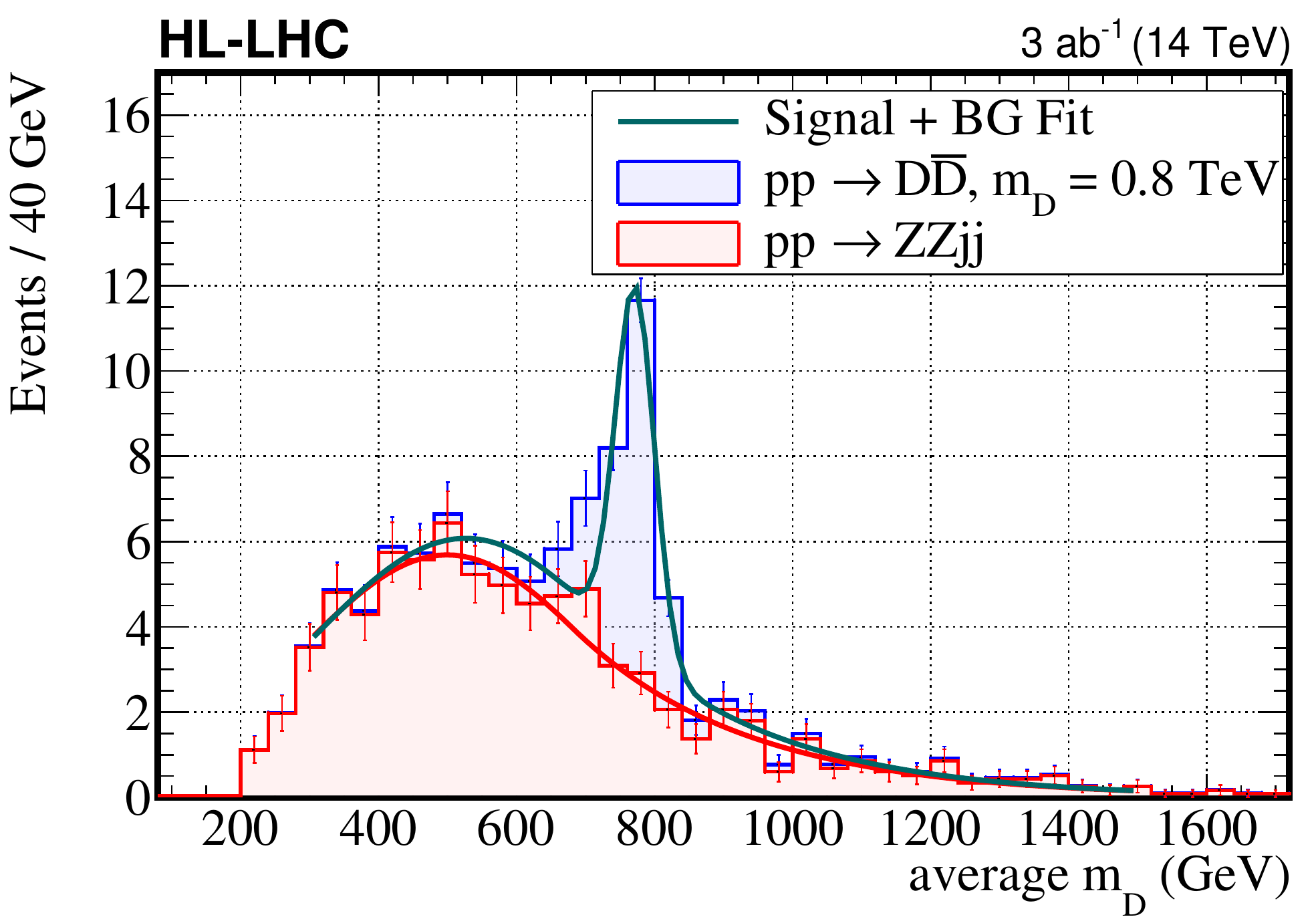}
\label{fig:5}       
  \includegraphics[width=.9\linewidth]{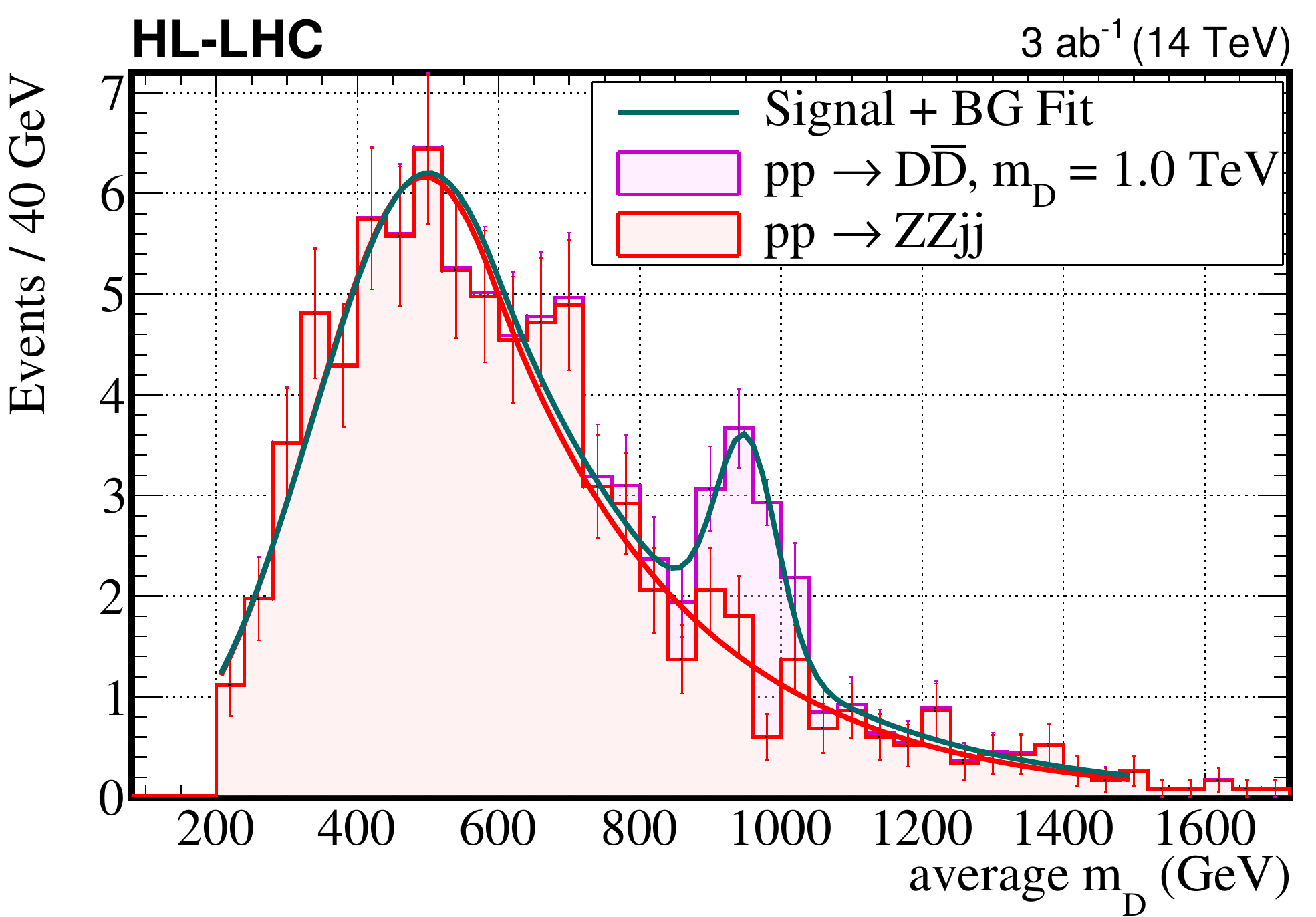}
\caption{Distribution of average reconstructed $D$ quark invariant mass $(m_{D_1} + m_{D_2})/2$ for HL-LHC conditions for background and signals with $m_D = 600$~GeV (top), $800$~GeV (middle) and $1000$~GeV (bottom). Results of the fit to the sum of a Gaussian and Crystal Ball functions are also shown.}
\label{fig:mDaveHLLHC}
\end{figure}

\begin{figure}[!h]
\centering
  \includegraphics[width=.9\linewidth]{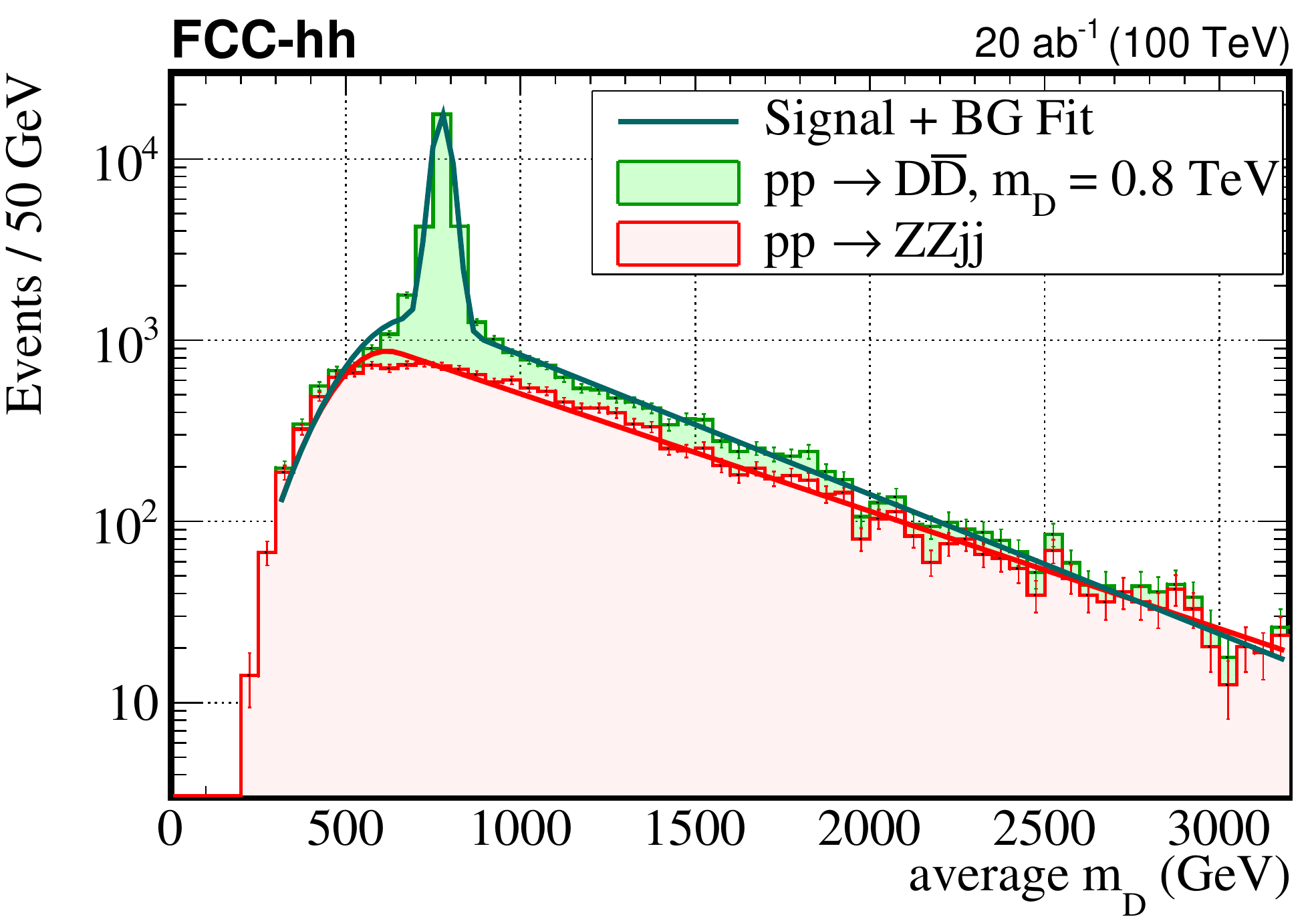}
\label{fig:4}       
  \includegraphics[width=.9\linewidth]{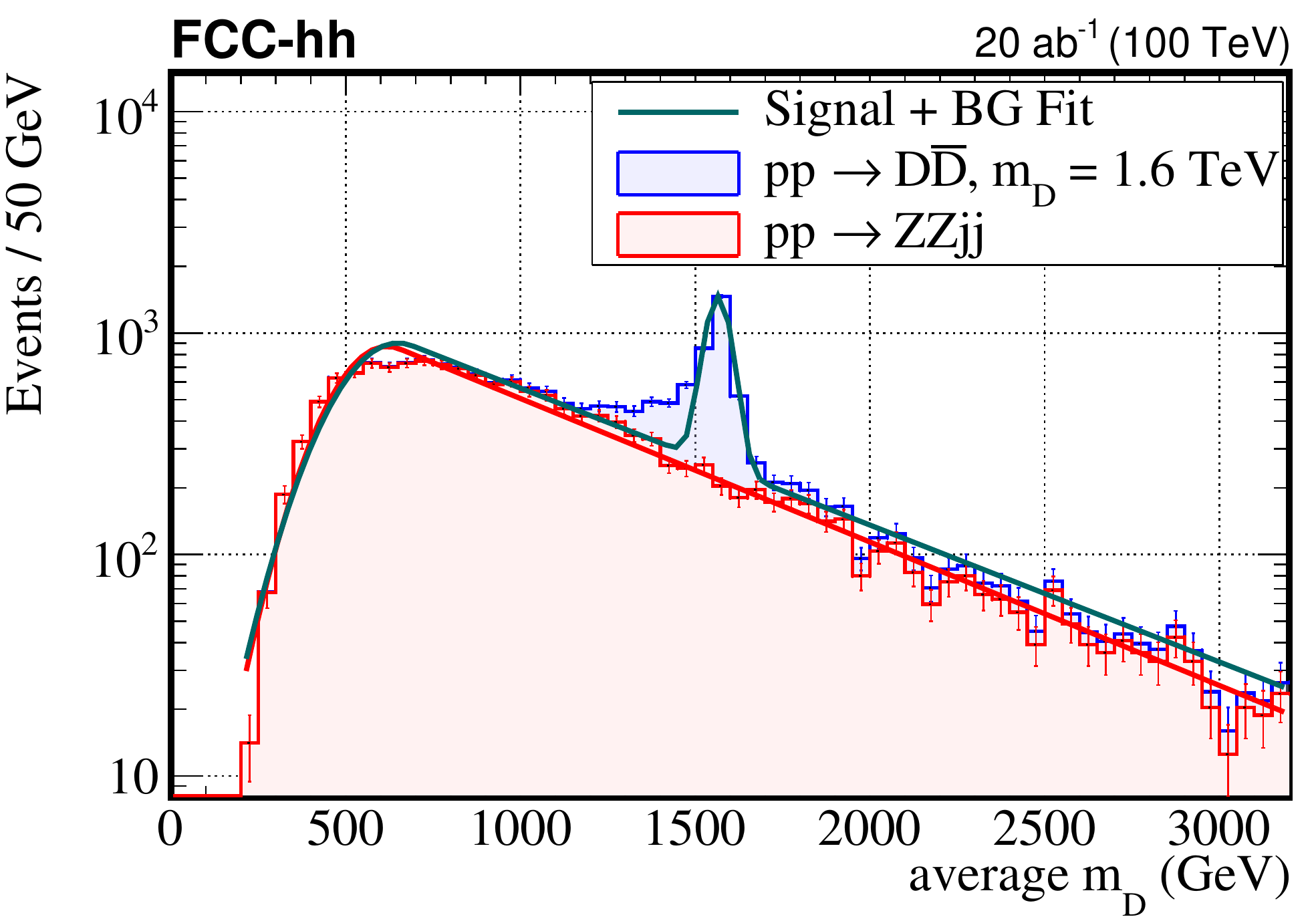}
\label{fig:5}  
  \includegraphics[width=.9\linewidth]{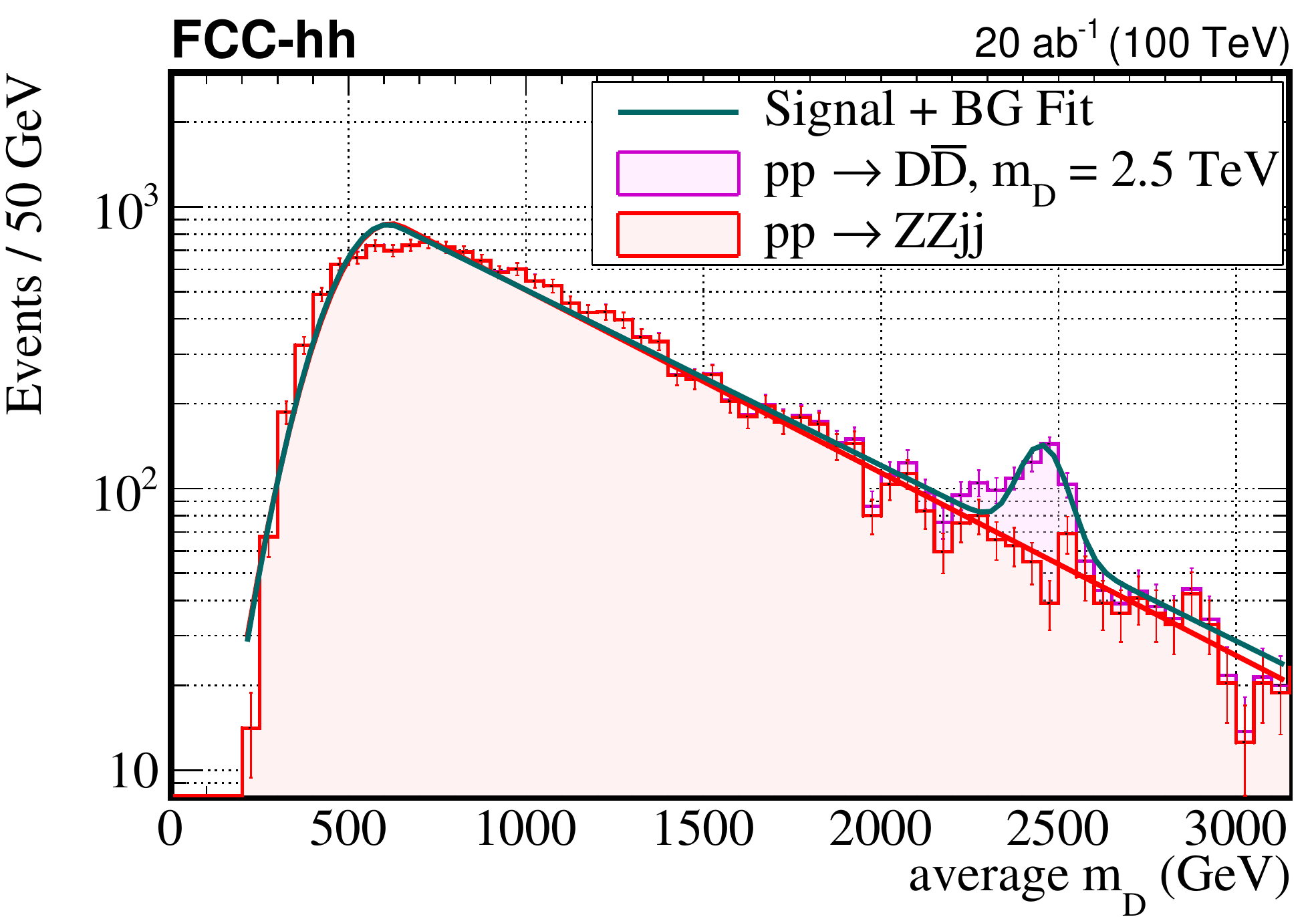}
\caption{Distribution of average reconstructed $D$ quark invariant mass $(m_{D_1} + m_{D_2})/2$ for FCC-hh conditions for background and signals with $m_D = 800$~GeV (top), $1600$~GeV (middle) and $2500$~GeV (bottom). Results of the fit to the sum of a Gaussian and Crystal Ball functions are also shown.}
\label{fig:mDaveFCC}
\end{figure}

The distribution of the average reconstructed $D$ quark invariant mass $(m_{D_1} + m_{D_2})/2$ in the signal and background events that remain after selection are shown in Figures~\ref{fig:mDaveHLLHC} and \ref{fig:mDaveFCC} for different generated $D$ quark masses for HL-LHC and FCC-hh, respectively.  Signal events are seen to peak visibly over the falling background distributions. { In order to reduce the statistical fluctuations due to limited amount of statistics,} the signal and background  distributions can be modelled with a Gaussian function and a Crystal Ball function, respectively. 
{
The signal and background yields are obtained from the total events distribution, by fitting it to the sum of these two functions. 
}The initial fit parameters for the Crystal Ball and Gaussian functions were determined by performing independent fits to the signal and background distributions. The resulting fits are also shown in the same figures.


\begin{figure}[!h]
  \includegraphics[width=.9\linewidth]{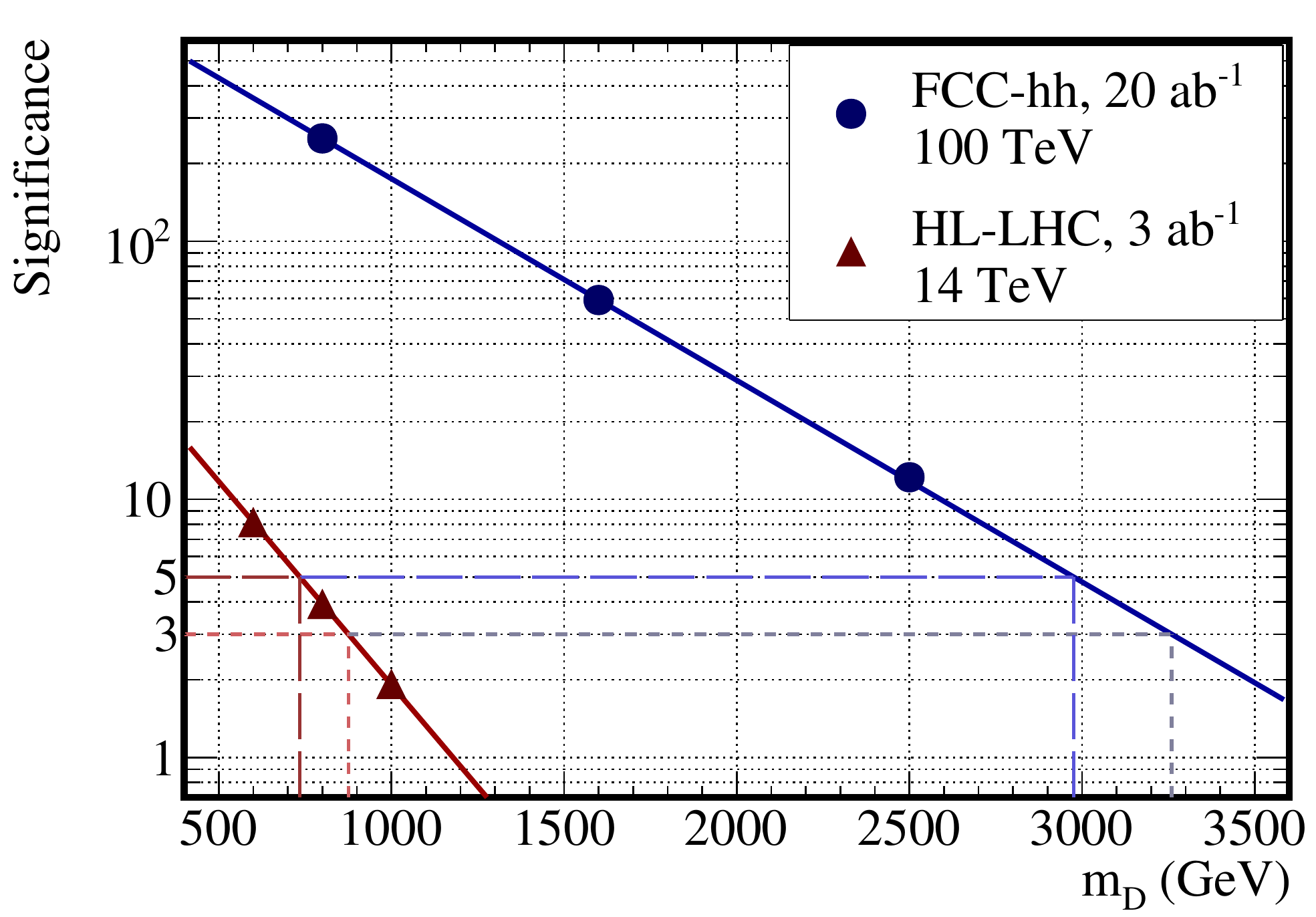}
\caption{Signal significance as a function of $D$ quark mass for HL-LHC and FCC-hh.}
\label{fig:significance}
\end{figure}

\begin{figure}[!h]
 
  \includegraphics[width=.9\linewidth]{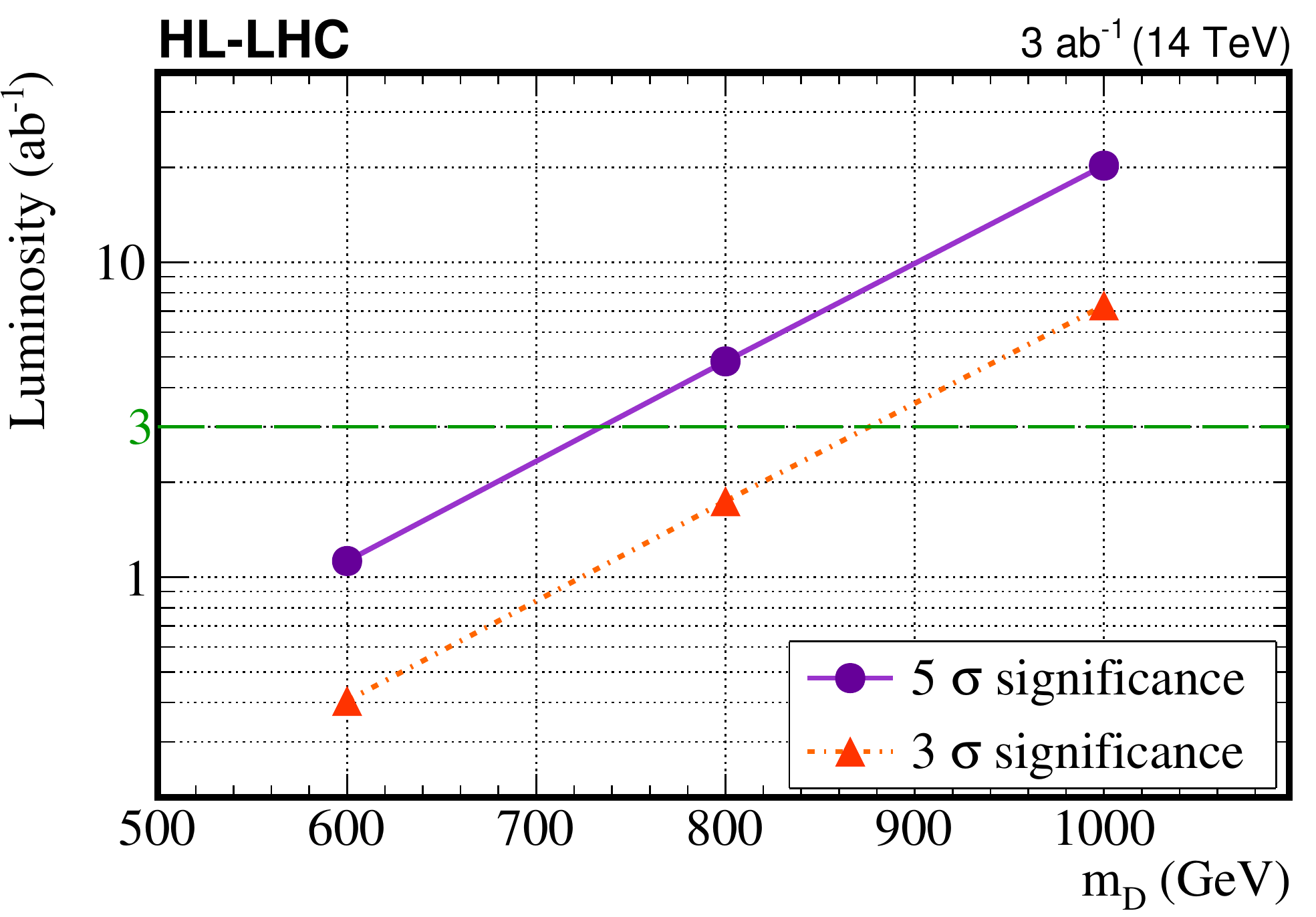}  
  \includegraphics[width=0.9\linewidth]{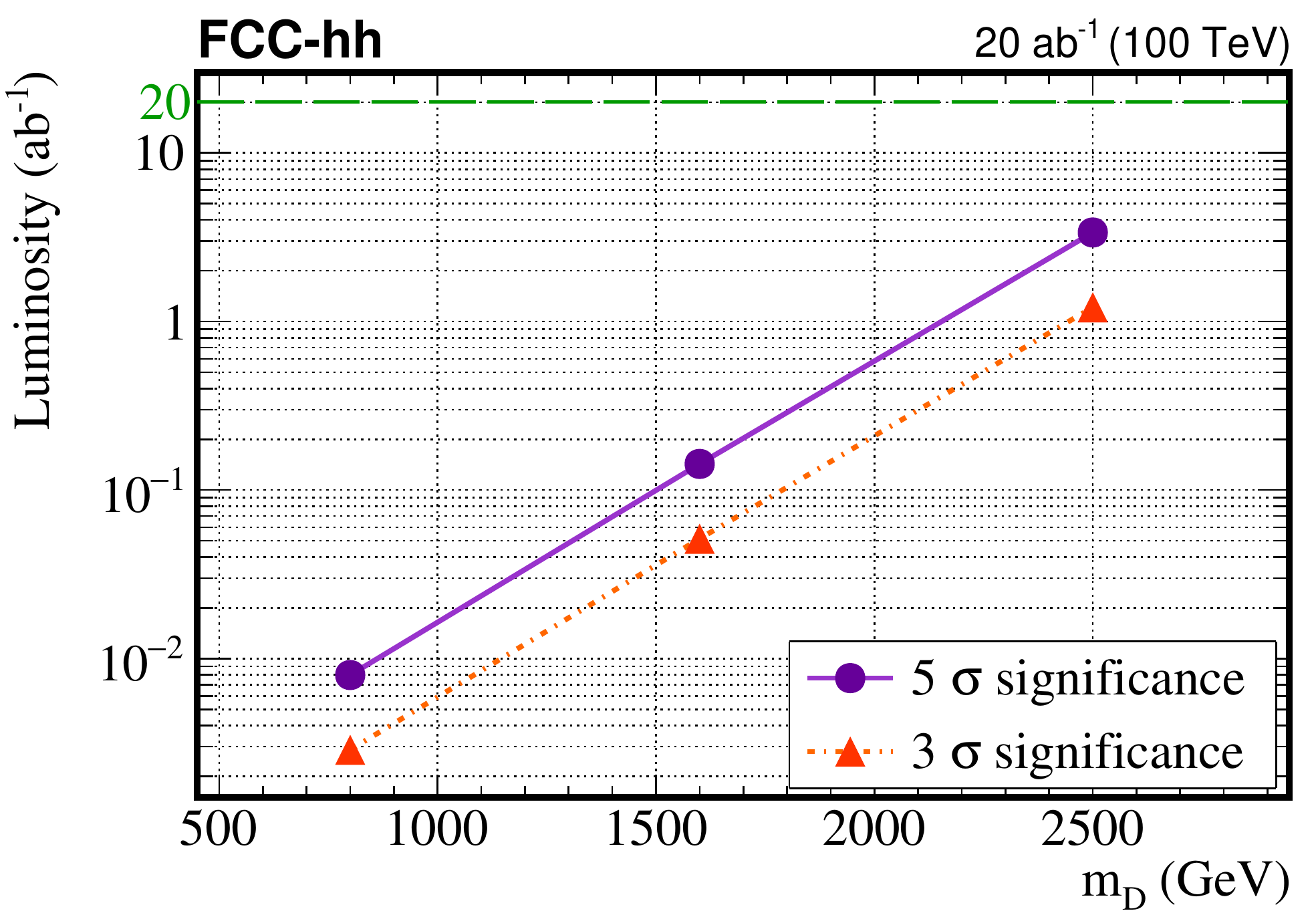}    
  \label{luminosityHLLHC}
\caption{The integrated luminosity needed for $3\sigma$ observation and $5\sigma$ discovery as a function of $D$ quark mass for HL-LHC (top) and FCC-hh (bottom).}
\label{fig:luminosity}
\end{figure}

The fit results are then used for estimating the final signal and background yields denoted as $S$ and $B$.  These are obtained by integrating the fitted Gaussian and Crystal Ball functions in a range defined by two standard deviations mass window around the Gaussian mean.  The obtained values for each $D$ quark mass are then used for calculating the signal significance $\sigma_{DD}$ defined as: 
\begin{align}
    \sigma_{DD} &\equiv \sqrt{2\times\left[\left(S+B\right) \ln{\left(1+\frac{S}{B}\right)} -S \right]} \; .
\end{align} 
The yields $S$ and $B$ along with the significance obtained for each simulated mass point are shown in Tables~\ref{tab:sigHLLHC} and~\ref{tab:sigFCC} for HL-LHC and FCC-hh, respectively.  Signal significance values are also shown in Figure~\ref{fig:significance}, plotted against the $D$ quark mass. A linear function is fitted to the plot to estimate the dependence of significance on $D$ quark mass.  The $D$ quark mass values, for which it would be possible to make an observation ($3\sigma$) or a discovery ($5\sigma$),  are then calculated from the linear function obtained from the fit, and are shown in {Table~\ref{tab:discoveryreach}} for HL-LHC and FCC-hh.  Finally, the integrated luminosities required for $3\sigma$ observation and $5\sigma$ discovery at HL-LHC and FCC-hh are plotted versus $D$ quark mass in Figure~\ref{fig:luminosity}.

\begin{table}[!h]
\centering
\caption{Signal and background yields and significance for different $D$ quark masses at HL-LHC.}
\label{tab:sigHLLHC}       
\begin{tabular}{lcccc}
\hline
D quark mass & S & B & $\sigma$\\ \hline
600 GeV & 46 & 19 & 8.2 \\
800 GeV & 15 & 10 & 3.9 \\
1000 GeV & 6 & 7 & 1.9 \\
\hline
\end{tabular}
\end{table}
\begin{table}[!h]
\centering
\caption{Signal and background yields and significance for different $D$ quark masses at FCC-hh.}
\label{tab:sigFCC}       
\begin{tabular}{lcccc}
\hline
D quark mass & S & B & $\sigma$\\ \hline
800 GeV & 21359 & 2690 & 250 \\
1600 GeV & 2123 & 740 & 59 \\
2500 GeV & 241 & 318 & 12 \\
\hline
\end{tabular}
\end{table}

\begin{table}[!h]
\centering
\caption{Upper limit on $D$ quark masses for $3\sigma$ observation and $5\sigma$ discovery for HL-LHC and FCC-hh.}
\label{tab:discoveryreach}       
\begin{tabular}{lccc}
\hline
Experiment & $3\sigma$ observation &  $5\sigma$ discovery \\ \hline
HL-LHC & 880 GeV & 730 GeV \\
FCC-hh & 3260 GeV & 2980 GeV \\
\hline
\end{tabular}
\end{table}

\section{Conclusions}
\label{sec:conclusions}

In this paper, we studied the feasibility of discovering pair-produced down type iso-singlet quarks $D$ at the High Luminosity LHC and the hadronic scenario for the Future Circular Collider.  The search was designed in the $4\ell + 2j$ channel, targeting the $D \rightarrow Zd \rightarrow \ell^+\ell^- d$ decay mode, which is not accessible at the LHC. Despite its relative low sensitivity, this channel is expected to provide the most precise reconstruction of the $D$ quark mass. Furthermore, in case of $D$ quark discovery through a higher sensitivity channel, $ZZ \rightarrow 4\ell$ channel would help to estimate relative branching ratios, thus leading to a preliminary understanding of the underlying model properties. However, extracting further information on the model would require observing and measuring iso-singlet partners of different quark types.

The analysis consisted of a basic event selection followed by a two-step reconstruction of the $D$ quark masses, where the $Z$ bosons were reconstructed in the first step.  A $\chi^2$ optimization was used for finding the combination giving the best $D$ quark candidates. A further selection was applied on the $\chi^2$ to discriminate signal events from the background.  Finally, a fit was performed on the average $D$ quark invariant mass distribution to obtain event yields and sensitivity.

The study showed that the $5\sigma$ discovery reach for $D$ quark mass at HL-LHC is possible, and is around 730 GeV for the full run period, while FCC-hh can reach up to 2980 GeV, considering only the $4\ell + 2j$ decay channel. It also demonstrated that FCC-hh requires about two orders of magnitude less integrated luminosity than HL-LHC for discovering $D$ quarks at a given mass. Therefore searches for $E_6$ GUT models using $4\ell + 2j$ channel would benefit from FCC-hh. { Sensitivity of FCC-hh could further be enhanced by the addition of final states with boosted Z bosons decaying to boosted collimated lepton jets.}

As a side note, this study showed an example of how extensively the analysis description language (ADL) concept and its runtime interpreter implementation, CutLang, can be used to benefit particle physics analyses. This approach allows performing the analysis algorithm steps (e.g. object definitions, object reconstructions, histogramming) in an  easy and descriptive way. 

\begin{acknowledgements}
SS is supported by the National Research Foundation of Korea (NRF), funded by the Ministry of Science \& ICT under contract NRF-2008-00460.  This study started during CERN Summer Student Program 2019, in which AP was a participant. GU would like to dedicate this paper to his parents Turkan and Olgun Unel (who both passed away during its preparation), for nurturing him into becoming the person he is.
\end{acknowledgements}

\setcounter{section}{0}
\renewcommand{\thesection}{\Alph{section}}
\section{Appendix: Explanation of the ADL implementation  
\label{sec:appendix}}
We provide details on the ADL/CutLang implementation of the analysis in this paper, in particular with the aim to clarify the implementation of composite object reconstruction and optimization. 
\subsection{Object selection}
Electron, muon and jet selection based on object properties, e.g. transverse momentum and pseudorapidity, is expressed as :
\renewcommand{\labelitemi}{$\bullet$}

{\begin{verbatim}
object goodJet
  take JET
  select Pt(JET) > 50 
  select abs(Eta(JET)) < 4 

object goodEle
  take ELE
  select Pt(ELE) > 20 
  select abs(Eta(ELE)) < 4 

object goodMuo
  take MUO
  select Pt(MUO) > 20 
  select abs(Eta(MUO)) < 4 
\end{verbatim}}
where \verb!JET!, \verb!ELE!, \verb!MUO! are the original objects from the input event files and \verb!goodJet!, \verb!goodEle! and \verb!goodMuo! are the derived objects.

Selected electrons and muons can be combined to define the unified set of leptons as 

\begin{verbatim}
object goodLep  : Union (goodEle, goodMuo)
\end{verbatim}

\subsection{Definitions}

ADL allows to define aliases for event variables or reconstructed particles through the usage of the \verb!define! keyword.  Shorthand notations for reconstructed $Z$ bosons and $D$ quarks, optimization criteria and selection variables are given in this section of the ADL file.

\subsection{Event selection}

Event selections in ADL are described within \verb!region! blocks defined for each selection region.  This analysis has a single search channel, which is described in the \verb!DDselection! region block, which starts by selecting the analysis final state of $4\ell+2j$ as:

\begin{verbatim}
region DDselection
  select Size(goodLep)  >= 4
  select Size(goodJet)  >= 2
\end{verbatim}
Subsequent optimization and selection requirements are also given in this region as described below.

\subsection{Z and D reconstruction}

In this analysis, the particles for reconstructing $Z$ and $D$ must be combined such that the resulting $Z$ and $D$ would best satisfy the criteria defined by an optimization rule. Therefore the indices of the particles combined are not known before the optimization, and are only determined after the optimization.  In CutLang, negative numbers are used for specifying such indices of particles that would be combined through a $\chi^2$ optimization.  Following this approach, $Z$ reconstruction is written as 
\begin{verbatim}
define Zreco1 = goodLep[-1] goodLep[-1]
define Zreco2 = goodLep[-3] goodLep[-3]
\end{verbatim}
Here the lepton indices are to be determined at run time for each event according to an optimization rule, yet to be defined. The repeated indices stress that in combining two leptons to reconstruct a $Z$ boson, the order is unimportant. The \verb|chi2ZZ| variable to be minimized in order to reconstruct the two $Z$ candidates is defined as 
{\begin{verbatim}
define chi2ZZ = (mZ1 - 91.2)^2 
              + (mZ2 - 91.2)^2 
              + (999*{Zreco1}pdgID)^2 
              + (999*{Zreco2}pdgID)^2
\end{verbatim}}
Given a variable $x$ with an optimal value $v$, the operator \verb!~=! is used to calculate the particle combination that gives an $x$ value \textbf{closest to} $v$.
The optimization criteria \verb!chi2ZZ! is finally called after the initial event selection, with the syntax 
{\begin{verbatim}
select chi2ZZ ~= 0
\end{verbatim}}

Note that, CutLang takes the PDG ID of a reconstructed object to be the sum of the PDG ID of its constituent objects. As the constituents of \verb!Zreco! must be a lepton-antilepton pair, \texttt{pdgID} of the \verb!Zreco! itself has to be zero. Here, $999$ is a high enough weight factor to ensure flavour neutrality of the $Z$ boson candidates.  A further requirement of $Z$ boson charge to be $0$ is also applied as:
{\begin{verbatim}
select q(Zreco1) == 0
select q(Zreco2) == 0
\end{verbatim}}

Next, $D$ quark candidates are reconstructed using the previously obtained $Z$ bosons and jets.  As in the case for $Z$ bosons, the indices of the optimal jets cannot be known, therefore, are written as negative indices.  
\begin{verbatim}
define dj1 = goodJet[-2]
define dj2 = goodJet[-4]
define Dreco1 = Zreco1 dj1
define Dreco2 = Zreco2 dj2 
\end{verbatim}
All these expressions are then used for defining terms in the optimization condition for $D$ reconstruction:
\begin{verbatim}
define chi2mD  = ((mD1 - mD2)/mD)^2
\end{verbatim}
\begin{verbatim}
define  chi2PTj 
    = Hstep(PTjcut -PTj1)*(PTjcut/PTj1 - 1.0)
    + Hstep(PTjcut - PTj2)*(PTjcut/PTj2 - 1.0)   
\end{verbatim}
\begin{verbatim}
define  chidRDD = (dRDD/3.14 - 1.0)^2 
\end{verbatim}

These terms are added to obtain $\chi^2_{DD}$ 
\begin{verbatim}
define chi2DD  = chimD + chiPTj + chidRDD   
\end{verbatim}
Finally, a selection criteria is applied on $\chi^2_{DD}$ in the \verb!DDselection! region as:
\begin{verbatim}
select chi2DD < chi2DDcut
\end{verbatim}

\subsection{Histogramming}

CutLang is designed to be a complete tool for event processing and visualization tasks in an analysis, and therefore allows to define and fill histograms at runtime. The CutLang syntax to plot 1D histograms of a variable is given below :
\begin{verbatim}
histo [label], "[title]", [no. of bins],
      [lower limit], [upper limit], [variable]
\end{verbatim}
Following are the definitions of histograms which were eventually plotted in Figures~\ref{fig:mZ},~\ref{fig:chi2DD},~\ref{fig:mDaveHLLHC} and~\ref{fig:mDaveFCC}:
\begin{verbatim}
histo hmZ1, "Z candidate1 mass (GeV)", 320,
                                 0, 3200, mZ1
histo hmZ2, "Z candidate2 mass (GeV)", 320,
                                 0, 3200, mZ2
histo hchi2DD, "chi2DD ", 200, 0, 10, chi2DD
histo hmD, "D candidate mass (GeV)", 320,
                                 0, 3200, mD
\end{verbatim}

\bibliographystyle{spphys} 
\bibliography{main}
\end{document}